\documentclass[aps,floatfix]{revtex4}

\usepackage{amssymb,hyperref,hypcap,
ae,tikz,color}
\usepackage{latexsym}
\usepackage{graphicx}
\usepackage{xcolor}

\usepackage[T1]{fontenc}
\usepackage[latin9]{inputenc}

\usepackage{mathrsfs}
\usepackage{amsmath}
\usepackage{amssymb}
\usepackage{esint}
\usepackage{babel}

\hypersetup{
	colorlinks=true,
	citecolor=green,
    linkcolor=blue,
	}

\newcommand{\be}{\begin{equation}}
\newcommand{\ee}{\end{equation}}
\newcommand{\bea}{\begin{eqnarray}}
\newcommand{\eea}{\end{eqnarray}}

\begin{document}

\title{Dynamical conditions and causal transport of dissipative spherical collapse in $f(R,T)$ gravity}

\author{Sarbari Guha and Uttaran Ghosh}
\affiliation{Department of Physics, St. Xavier's College (Autonomous), Kolkata 700016, India}

\begin{abstract}
In this paper, we have investigated the non-adiabatic spherical gravitational collapse in the framework of the $f(R,T)$ theory of gravity with a locally
anisotropic fluid that undergoes dissipation in the form of heat flux, free-streaming radiation, and shearing viscosity. The dynamical equations are analyzed in detail, both in the Newtonian and post-Newtonian regimes. Finally we couple the dynamical equations to the full causal transport equation in the context of Israel-Stewart theory of dissipative systems. This yields us a better understanding of the collapse dynamics and may be connected to various astrophysical consequences.
\end{abstract}

\maketitle

\section{Introduction}

The final outcome of stellar evolution depends not only on the size of the object undergoing collapse \cite{Chandra} but also on other physical parameters \cite{8}. On account of the high energy dissipation during collapse, more massive stars
tend to be more unstable because of massive radiation transport and rapid loss of nuclear fuel \cite{8,9}. The effects of dissipation and slight change in isotropy alters the subsequent evolution of the collapsing system considerably. Several researchers \cite{5,10,11,12,13,14,15,16,Santos,OSK,OPS} have studied extensively the phenomenon of gravitational collapse of fluids in presence or absence of various conditions like anisotropy, radiation, dissipation, expansion, shear, along with the conditions of dynamical instability and causal transport phenomenon occurring during gravitational collapse, and the list is even longer. The subject of spherical gravitational collapse have always been at the heart of astrophysical investigations on account of the symmetry of the matter distribution, leading to simplification of the field equations, and also due to the fact that some classes of realistic gravitational collapse can be modelled as spherical collapse with only small deviations. The analysis is also simplified due to the absence of gravitational waves. Studies of radiating spherical collapse in presence of heat flow \cite{GGM,MGG1}, relaxational effects in stellar heat transport during collapse \cite{GMM} and temperature profile of such a collapse within the framework of causal thermodynamics have been conducted \cite{MGG2}. Some authors \cite{GB,PHDMS} have also studied the dissipative collapse of charged configurations.

The limitations of General Relativity (GR) on large scales led to the investigation of astrophysical processes in modified theories of gravity, e.g., theories which provide improved models of the accelerating universe. Among these modified theories (some of the review articles are listed in  \cite{17a}), the $f(R)$ gravity presents a very elementary modification by including higher order curvature terms to incorporate the dark energy components, as well as the inflationary phase \cite{DEinfl}. Not only does $f(R)$ gravity reproduce the $\Lambda$CDM epoch, or is able to mimic the cosmological constant at the present epoch, but it can also unify the entire evolution history of the universe \cite{evolution}. The null dust non-static exact solutions in $f(R)$ gravity constrained by constant curvature describing anti de-Sitter background evolution, was studied by Ghosh and Maharaj \cite{18}. Cembranos et al. \cite{19} studied the evolution of gravitating sources in the presence of dust fluid in a general $f(R)$ model with a view to determine the possible constraints on such models. Goswami et. al. studied the collapse of massive stars in $f(R)$ gravity and showed that the extra matching conditions arising in modified gravity impose strong constraints on the stellar structure and thermodynamic properties \cite{Goswami}. Chakrabarti and Banerjee investigated the collapse of a perfect fluid source described by Lemaitre-Tolman-Bondi type geometry \cite{CB}. Sharif and Yousaf studied the dynamical instability of charged spherical collapse in expansion-free condition \cite{SY2}, and the stability of dissipative charged spherical collapse in the CDTT-$f(R)$ model \cite{SY3}.

In 2011, Harko et al. \cite{22} introduced the $f(R,T)$ theory of gravity, based on the non-minimal matter-to-geometry coupling considered by earlier workers \cite{22a,22b,22c}. The action in $f(R,T)$ gravity includes an arbitrary function of the  Ricciscalar $R$ and the trace $T$ of the energy-momentum tensor. The inclusion of $T$ takes care of quantum effects or the existence of exotic matter. The higher-curvature terms in this theory can address the flatness problem which can be seen in the galactic rotation curves. A combination of a term quadratic in $R$ (as in the Starobinsky model \cite{Starobinsky}) along with linear terms in $R$ and $T$ as used in \cite{SZ2}, provides a useful $f(R,T)$ function. It was realized \cite{alvarenga2013} that a strong coupling of the curvature $R$ with the trace $T$ leads to the violation of the conventional equation of continuity, as a result of which, an extra force arises in the geodesic equation and test particles do not follow a geodesic motion. However, this problem can be resolved by choosing a suitable $f(R,T)$ function. These features of $f(R,T)$ gravity make it \emph{a candidate suitable for} the investigation of various astrophysical phenomena within the context of this theory. The astrophysical, cosmological and thermodynamic implications of $f(R,T)$ gravity have been widely studied by several groups \cite{alvarenga2013,barrientos2014,Faraoni2009,23,24,SZ1}. Houndjo \cite{Houndjo} demonstrated a possible unification of the accelerated expansion phase with the matter dominated era in cosmic evolution by choosing a viable $f(R,T)$ function subject to appropriate constraints, which ensure that either no ghost state arises where dark energy becomes responsible for the accelerated expansion, or no tachyon arises. Subsequently Odintsov and S\'{a}ez-G\'{o}mez \cite{OSG} proposed the $f(R,T,R_{\mu\nu}T^{\mu\nu})$ theory as an extension of $f(R,T)$ gravity. They derived the general FRLW field equations in the presence of the $R_{\mu\nu}T^{\mu\nu}$ coupling terms and studied several cosmological solutions, thereby establishing that the matter sector behaves differently from that in GR, because the continuity equations are different. In this theory too, the problem can be dealt with by considering an appropriate $f(R,T,R_{\mu\nu}T^{\mu\nu})$ function, so that one can assume the usual evolution for a perfect fluid matter even in presence of the extra degree of freedom.

Sharif and Yousaf \cite{SY} considered the dynamical analysis in $f(R,T)$ gravity with non-null expansion scalar at Newtonian and post-Newtonian eras and examined the role played by matter variables on stellar structure. Others investigated the dynamical instability of spherical stars in locally anisotropic environment in $f(R,T)$ theory under various situations like expansion-free, shear-free conditions and from other perspectives \cite{NZ1,NZ2,NZ3}. Yousaf et al. \cite{YBB} studied the factors leading to irregularities in the case of a spherically symmetric self-gravitating star consisting of dissipative matter and radiation in $f(R,T)$ gravity. In another work \cite{InflModGrav} they discussed the advantages of working with modified gravity. If the curvature is low, one can observe accelerated expansion of the universe, while for high value of curvature, the singularities can be made smoother. They \cite{YBBG} also studied the effect of charge on spherically symmetric gravitational vacuum stars (gravastars) in $f(R,T)$ gravity. Yousaf \cite{Y2020} considered a charged cylindrically symmetric gravastar with perfect fluid matter content in $f(R, T)$ gravity. It was found that
for an increase in the amount of charge in the gravastar, the proper length of its middle thin shell reduces. Also, the energy content
in the gravastar reduces for a rise in the amount of electric charge. The entropy of the system increases if the charge decreases or with an increase in the thickness of the shell. Bhatti et al. \cite{BYR} considered a spherically symmetric model of a gravastar and studied some of its characteristics in $f(R,G)$ theory, where $G$ is the Gauss-Bonnet term.

The physical behavior, stability and validity of energy conditions of compact objects under the effect of charge was examined in \cite{SS}, and the collapsing and expanding solutions of anisotropic charged cylinder was determined in \cite{SS1}. Some authors have studied the higher-dimensional perfect fluid collapse in $f(R,T)$ theory \cite{hdimPfcoll}. The effects of charge on the stability range of anisotropic spherical stellar model in the framework of $f(R,T)$ gravity have been studied in \cite{SW}. In the paper \cite{AA1} the exact solutions for non-static anisotropic self-gravitating source in $f(R,T)$ were derived, and the nature of singularity and trapped surfaces during the collapse of a charged perfect fluid was determined in \cite{AA}. They also studied the thermodynamic aspects of a viscous dissipative collapse in $f(R,T)$ gravity \cite{AA2}. Recently the complexity factor of a self-gravitating system was determined through the orthogonal splitting of the Riemann tensor and the behaviour of the complexity for a cylindrically symmetric system in $f(R,T)$ theory of gravity was studied \cite{ZA}.

A spherically symmetric perfect fluid model for a gravastar
in $f(R,T,Q)$ theory was studied in \cite{YBA}, where $Q = R_{\mu\nu}T^{\mu\nu}$. The paper \cite{YBF} studied
the instability criteria by the method of dynamical analysis for a
cylindrically symmetric, anisotropic radiating matter fluid in $f(R,T,Q)$
theory. The linear perturbation scheme was invoked under the assumption that the fluid is initially in hydrostatic equilibrium,
with dependence only on the radial coordinate, but later it begins
to undergo radial oscillations. The instability ranges were calculated in the Newtonian and post-Newtonian regimes, utilising the Harrison-Wheeler
equation of state. The same authors \cite{YBF2} discussed the instability ranges of compact stars by the method of dynamical analysis
in $f(R,T,Q)$ theory and adopting the linear perturbative scheme.
The energy-momentum tensor was that of an anisotropic perfect
fluid, and the interior metric was the most general spherically symmetric
with non-zero expansion scalar. Yousaf, Bamba, Bhatti and Farwa \cite{YBBF}, studied the dynamical evolution of a spherical anisotropic fluid
with non-vanishing expansion scalar using perturbative scheme, and its instability ranges calculated at the Newtonian and
post-Newtonian regimes. 

Gravitational collapse ensues only when a self-gravitating object is perturbed from its initial hydrostatic
equilibrium condition, when the internal pressure cannot counteract the gravitational contraction. This leads to the formation of compact objects like white dwarfs, neutron stars, black holes, etc. The effect of various factors in the formation of these objects can hardly be undermined. In this context, Yousaf \cite{Yousaf2018} explored the behavior of scalar variables associated with the shearing viscous dissipative anisotropic spherical stars in the framework of $f(G,T)$ theory. A correspondence was drawn between the metric scale factors, tidal forces and structure parameters of the collapsing configuration. As the technique of orthogonal breaking of the Riemann tensor helps us to study the reasons behind the emergence of inhomogeneities in the initially regular spheres as the collapse progresses, this technique was used to determine the modified structure scalars, and the role of these invariants in the evolutionary properties of radiating spheres, was explored. The analysis showed that the evolutionary phases of the spherical interiors can be well understood in terms of extended versions of scalar variables. The shear evolution equation (SEE), the expansion evolution equation (EEE) --- also known as the Raychaudhuri equation, and the Weyl differential equation (WDE) can be expressed with the help of these structure scalars. For constant
$G$ and $T$, in case of a dust ball, the cause of inhomogeneous energy density in the compact object which is initally regular, can be
expressed through the WDE using one of the structure scalars \cite{Yousaf2019a}. Depending on
the mass of the collapsing object \cite{Yousaf2019b, Yousaf2021}, the collapse can result in formation
of structures like black holes, neutron stars or white dwarfs.

It was Misner and Sharp \cite{MS} and Misner \cite{Misner}, who provided a complete description of the dynamical equations for
adiabatic and dissipative collapse. In fact, all the physical parameters associated with the collapsing matter, should obey the transport equations of causal thermodynamics. Earlier Tolman \cite{Tolman} had suggested that the heat energy associated with the collapsing matter should have an inertia associated with it, and this inertial term appears in the transport equations derived by Eckart and Landau-Lifshitz \cite{Eckart, Landau}. However the Landau-Eckart prescription could only be solved by a hyperbolic theory involving second-order entropy terms, as done by Muller, Israel and Stewart \cite{Muller, Israel1, Israel2, Israel3}. The dynamical equation can be recasted into the form of Newton's Second Law of Motion. The transport equation can be expressed in terms of the acceleration of the collapsing system, and after we couple the transport equation with the dynamical equation, both the inertial mass, and the gravitational mass are found to get reduced by the same factor, in agreement with the principle of equivalence. The thermal dissipative effects reduce the effective inertial mass term, and the inertial mass appears to lag behind the collapsing matter as the collapse progresses \cite{HerreraInertia}.

Herrera and Santos \cite{12} discussed the importance of dissipative processes in gravitational collapse and extended the Misner dynamical equations to include dissipation in the form of radial heat flow along with pure radiation in order to couple the dynamical equations with the transport equations. Herrera et al. \cite{PHDMS} applied the same prescription to charged, dissipative, spherically symmetric system with shear viscosity. The paper \cite{HPFT2009} examined spherically symmetric collapse with heat flow, radiation, shear and bulk viscosity. They applied a full causal treatment to all dissipative variables, and further included the viscous or heat coupling coefficients, which are known to influence the evolution of neutron stars \cite{Maartens}. Plane symmetric gravitational collapse in the Misner and Sharp formalism was presented for a fluid undergoing dissipation in the form of heat flow, radiation, shear and bulk viscosity, and the dynamical equations were coupled with the causal transport equations \cite{srGRG10}. The gravitational collapse of cylindrical viscous heat conducting anisotropic fluid was investigated in \cite{CC}, and the transport equation was derived to determine its effect on the collapsing system. The effect of dissipation on the dynamics of charged bulk viscous collapsing cylindrical source with heat flux was studied in \cite{SA}. Hydrodynamic and thermodynamic analysis of gravitational collapse of a locally anisotropic shear-free fluid with dissipative flows was presented in \cite{PS}. The exact model satisfied all the energy conditions
throughout the interior of the star and for the entire duration of collapse process, and provided a physically viable temperature profile in the context of
causal thermodynamics.

Spherically symmetric collapse of isotropic non-viscous fluid satisfying the energy conditions together with matter violating the null energy conditions was investigated in the framework of $f(R)$ theory and effect of such matter on the coupling of dynamical and transport equations, was studied in \cite{SK}. Dissipative gravitational collapse in non-static cylindrical symmetric geometry was examined in \cite{AbbasNazar2018} by using Misner-Sharp formalism in the framework of metric $f(R)$ gravity, and the dynamical equation was coupled with the full causal transportation equations in the context of Israel-Stewart formalism. Static configurations and radial stability of compact stars have also been studied within the context of $f(R,T)$ gravity \cite{PJRA}. Recently, the dynamics of non-adiabatic charged spherical gravitational collapse of an anisotropic fluid with heat flux was analyzed in the framework of $f(R,T)$ gravity using the Misner-Sharp approach \cite{AA3}. However, the authors did not consider the effect of shear and free-streaming radiation, nor were the instability conditions analysed.

In this paper, we intend to explore the dynamical instability and the causal heat transport of an uncharged anisotropic fluid with shear viscosity in presence of free-streaming radiation, undergoing spherically symmetric collapse in $f(R,T)$ gravity. The manuscript is organized as follows: In Section II, we present the $f(R,T)$ formalism, followed by the choice of the interior metric and the corresponding field equations in Section III. The dynamical equations are laid down in Section IV, and the perturbation analysis along with the Newtonian and post-Newtonian approximation are presented in Section V. The transport equations are derived in Section VI. Finally, the summary of our findings and the final outlook are presented in Section VII.

\section{The $f(R,T)$ formalism}
The modified Einstein-Hilbert action in $f(R,T)$ gravity is given by \cite{22}
\begin{equation}\label{1}
S=\int ~d^{4}x \sqrt{-g} \Bigg(\frac{f(R, T)}{16\pi G} + \mathcal{L} _ {m} \Bigg),
\end{equation}
where $g$ is the determinant of the 4D metric $g_{\mu\nu}$, $R$ is the Ricciscalar, $T$ is the trace of the stress-energy tensor $T_{\mu\nu}$, $G$ is the Newton's gravitational constant, $\mathcal{L} _ {m}$ denotes the Lagrangian density for matter fields, and $f(R,T)$ is a suitable analytic function of $R$ and $T$ that describes the gravitational interaction. The corresponding field equations of $f(R,T)$ gravity has the form:
\begin{equation}\label{field}
R_{\mu\nu}f_R (R, T) - \frac{1}{2}g_{\mu\nu}f(R,T) + (g_{\mu\nu}\Box - \nabla_\mu\nabla_\nu)f_R (R,T)
=8\pi G T^{m}_{\mu\nu} - f_T (R, T) \left( T^{m}_{\mu\nu} + \Theta_{\mu\nu} \right),
\end{equation}
where $f_{R}(R,T)$ and $f_{T}(R,T)$ denote the derivatives of $f(R,T)$
with respect to $R$ and $T$ respectively, ${\Box}= g^{\mu\nu} {\nabla}_{\mu}{\nabla}_{\nu}$ is the d'Alembertian
operator, ${\nabla}_{\mu}$ is the covariant derivative associated with the Levi-Civita connection of the metric tensor, and
$\Theta_{{\mu}{\nu}}$ is defined by
\begin{equation}\label{Theta}
\Theta_{{\mu}{\nu}} = g^{\alpha{\beta}}\frac{{\delta}T_{{\alpha}{\beta}}}{{\delta}g^{\mu{\nu}}} = - 2T^m_{{\mu}{\nu}} + g_{\mu\nu}\mathcal{L}_m
- 2g^{\alpha\beta}\frac{\partial^2\mathcal{L}_m}{{\partial} g^{\mu\nu}{\partial}g^{\alpha\beta}}.
\end{equation}
The covariant divergence of eq.~(\ref{field}) gives~\cite{barrientos2014}
\begin{eqnarray}\label{divEMT}
\hspace{-0.5cm}\nabla^{\mu}T_{\mu\nu}^{m} &=& \frac{f_{T}(R,T)}{8\pi G - f_{T}(R,T)} \left[(T_{\mu\nu}^{m}+\Theta_{\mu\nu})\nabla^{\mu}\ln f_{T}(R,T) + \nabla^{\mu}\Theta_{\mu\nu} - \frac{1}{2}g_{\mu\nu}\nabla^{\mu}T \right].
\end{eqnarray}
Thus the conventional energy-momentum tensor is not conserved, unlike in GR, which means that particles do not follow \emph{geodesic paths} in pure gravitational fields in the framework of $f(R,T)$ gravity.

While writing out the field equations, various types of $\mathcal{L} _ {m}$ can be chosen to model different types of matter. For this anisotropic fluid, we choose $\mathcal{L}_m = - \rho$, where $\rho$ is the energy density of the matter distribution, so that (\ref{Theta}) reduces to the form
$$\Theta_{\mu\nu}=-2T^{m}_{\mu\nu} - \rho g_{\mu\nu}.$$
Consequently, (\ref{field}) becomes
\begin{eqnarray}\label{7}
R_{{\mu}{\nu}}f_{R}(R,T) - \frac{1}{2}g_{{\mu}{\nu}}f(R,T) + (g_{{\mu}{\nu}}{\Box} - {\nabla}_{\mu}{\nabla}_{\nu})f_{R}(R,T) = 8{\pi}GT_{{\mu}{\nu}}^{m} + f_T (R, T) \left( T^{m}_{\mu\nu} + \rho g_{{\mu}{\nu}} \right) .
\end{eqnarray}
Faraoni~\cite{Faraoni2009} showed that both ${\mathcal{L}}_m= p$ and  ${\mathcal{L}}_m= - \rho$, where $p$ is the isotropic pressure of matter, are equivalent when the fluid couples minimally to gravity. However, this freedom of choice is lost when the pressure is not isotropic.

Equation (\ref{7}) can be written in the form of an effective field equation as follows:
\begin{equation}\label{8}
G_{\mu\nu} = R_{{\mu}{\nu}}-\frac{1}{2}Rg_{{\mu}{\nu}} = 8{\pi}G_{eff}T_{{\mu}{\nu}}^{m} + T^D_{{\mu}{\nu}},
\end{equation}
where $$ G_{eff}=\frac{1}{f_{R}(R,T)}\left(G+\frac{f_{T}(R,T)}{8\pi}\right)$$
is the effective gravitational coupling in $f(R,T)$ gravity, and
\begin{equation}\label{9}
T^D_{{\mu}{\nu}} = \frac{1}{f_{R}(R,T)}\left[ \rho g_{\mu\nu}f_{T}(R,T) + \frac{1}{2} (f(R,T) - Rf_{R}(R,T)) g_{\mu\nu} + ({\nabla}_{\mu}{\nabla}_{\nu} - g_{{\mu}{\nu}}{\Box}) f_{R}(R,T)\right]
\end{equation}
represents the contribution to the energy-momentum tensor from the matter-curvature coupling and also includes non-equilibrium description of the field equations.

The full form of eq. (\ref{8}) in $f (R,T )$ gravity is then
\begin{equation}\label{EFTfull}
G_{\mu\nu}=T_{\mu\nu}^{eff}=\frac{1}{f_R}\left[(8\pi G + f_T)T^{m}_{\mu\nu} + \rho g_{\mu\nu}f_T + \frac{1}{2}(f - Rf_R)g_{\mu\nu} + (\nabla_\mu \nabla_\nu - g_{\mu\nu}\Box)f_R\right],
\end{equation}
where $T_{\mu\nu}^{eff}$ is the effective energy-momentum tensor incorporating gravitational effects. Our equation (\ref{EFTfull}) follows the convention adopted by several authors beginning with Harko and his collaborators (see for example \cite{22,22b,alvarenga2013,barrientos2014,SY,NZ1,NZ2,NZ3,SW}). However, instead of the term $(\nabla_\mu \nabla_\nu - g_{\mu\nu}\Box)f_R(R,T)$, Yousaf et. al \cite{YBB} considered the term $\left({\nabla}_\mu{\nabla}_\nu + g_{\mu\nu}{\Box}\right)f_R(R,T)$ in the $f(R,T)$ field equations (2) in their paper. This led to a difference in the various components of the field equations obtained by them. Here we have retained the expression given in \cite{22}.

In the subsequent analysis we will assume $8\pi G = 1$ for the sake of simplicity of presentation.

\section{Interior metric and Field equations for $f(R,T)$ gravity}

The physical system consists of a timelike 3D bounding surface, $\Sigma$, which divides the 4D spacetime into an interior and an exterior portion, denoted by $V^-$ and $V^+$ respectively. We assume that the region inside $\Sigma$ is modeled by the general non-static non-rotating spherically symmetric metric
\begin{equation}\label{intmetric}
ds^2_-=-A^2dt^{2}+B^2dr^{2}+C^2(d\theta^{2}+\sin^{2}\theta d\phi^{2}),
\end{equation}
where $A, B$ and $C$ are in general functions of both $t$ and $r$. The interior spacetime is filled with shearing viscous, locally anisotropic and radiating fluid described by the energy-momentum tensor $T^m_{\mu\nu}$:
\begin{equation}\label{3}
T^m_{\mu\nu}=(\rho + p_{\perp})V_{\mu}V_{\nu} + p_{\perp}g_{\mu\nu} + (p_r - p_{\perp})\chi_{\mu}\chi_{\nu} + q_{\mu}V_{\nu} + V_{\mu}q_{\nu}+{\epsilon}l_\mu l_\nu-2{\eta}{\sigma}_{\mu\nu},
\end{equation}
where $\rho$ is the energy density, $p_{\perp}$ the tangential pressure, $p_r$ the radial pressure, $q^{\mu}$ the heat flux, $\epsilon$ is the radiation density, $V^{\mu}$ is the time-like four velocity of the fluid, $\chi^{\mu}$ is a unit spatial vector along the radial direction, $l^{\mu}$ is a null 4-vector, and $\eta$ $(> 0)$ is the coefficient of shearing viscosity, respectively, such that
\begin{equation}
V^{\mu}V_{\mu}=-1, \;\; V^{\mu}q_{\mu}=0, \;\; \chi^{\mu}\chi_{\mu}=1, \;\; \chi^{\mu}V_{\mu}=0, \;\; l^{\mu} V_{\mu}=-1, \;\; l^{\mu}l_{\mu}=0.
\end{equation}
In comoving coordinates, we have
\begin{equation}\label{02}
V^{\mu}=A^{-1}\delta^{\mu}_{0},\quad
\chi^{\mu}=B^{-1}\delta^{\mu}_{1}, \quad
l^{\mu}=A^{-1}\delta^{\mu}_0+B^{-1}\delta^{\mu}_{1}, \quad
q^\mu = B^{-1}q{\delta}^{{\mu}}_{1} = B^{-1}(0,~q,~0,~0),
\end{equation}
where $q=q(r,t)$ and $\epsilon=\epsilon(r,t)$. The acceleration vector $a^{\mu}$, the expansion scalar $\Theta_1$ and the magnitude of the shear scalar
$\sigma$ are given by \cite{11,TRM}:
\begin{eqnarray}
  a^{\mu} &=& \left(0, \frac{A'}{AB^2},0,0\right), \label{08} \\
  \Theta_1 &=& \frac{1}{A}\left(\frac{\dot{B}}{B}+2\frac{\dot{C}}{C}\right),\label{09} \\
  \sigma &=& -\frac{1}{3A}\left(\frac{\dot{B}}{B}-\frac{\dot{C}}{C}\right),\label{10}
\end{eqnarray}
where primes and dots on the metric functions denote partial derivatives with respect to $r$ and $t$ respectively.

The components of the Einstein tensor for the interior metric are given by
\begin{eqnarray}
\label{eq:g6a}
G_{00} &=& A^2\left[ \frac{2}{A^2}\frac{\dot{B}}{B}\frac{\dot{C}}{C} + \frac{\dot{C}^2}{A^2C^2} + \frac{1}{C^2} -
\frac{1}{B^2}\left( \frac{2C''}{C} + \frac{{C'}^2}{C^2} - \frac{2B'}{B}\frac{C'}{C}\right)\right],\\
\label{eq:g6b} G_{11} &=& B^2 \left[\frac{1}{A^2} \left( - \frac{2 \ddot{C}}{C} - \frac{\dot{C}^2}{C^2} + \frac{2\dot{A}}{A}\frac{\dot{C}}{C}\right) +
\frac{1}{B^2} \left(\frac{{C'}^2}{C^2} + \frac{2 A'}{A}\frac{C'}{C}\right) - \frac{1}{C^2} \right],\\
\label{eq:g6c} G_{22} &=& C^2 \left[ - \frac{1}{A^2} \left(\frac{\ddot{B}}{B} - \frac{\dot{A}}{A}\frac{\dot{B}}{B} + \frac{\dot{B}}{B}\frac{\dot{C}}{C} -
\frac{\dot{A}}{A}\frac{\dot{C}}{C} + \frac{\ddot{C}}{C}\right) + \frac{1}{B^2} \left(\frac{A''}{A} - \frac{A'}{A}\frac{B'}{B}
+ \frac{A'}{A}\frac{C'}{C} - \frac{B'}{B}\frac{C'}{C} + \frac{C''}{C}\right) \right],\\
\label{eq:g6d} G_{01} &=& 2 \left(-\frac{\dot{C}'}{C} + \frac{A'}{A}\frac{\dot{C}}{C} + \frac{\dot{B}}{B}\frac{C'}{C}\right).
\end{eqnarray}

The Ricciscalar for the interior metric (\ref{intmetric}) is given by
\begin{equation}\label{ricciscalar_int}
R = - \frac{2}{A^2} \left( \frac{\dot{A}\dot{B}}{AB} + \frac{2\dot{A}\dot{C}}{AC} - \frac{\ddot{B}}{B} - \frac{2\dot{B}\dot{C}}{BC} - \frac{\dot{C}^2}{C^2} - \frac{2\ddot{C}}{C} \right) - \frac{2}{B^2} \left( \frac{A''}{A} - \frac{A'B'}{AB} + \frac{2A'C'}{AC} + \frac{C'^2}{C^2} - \frac{2B'C'}{BC} + \frac{2C''}{C} \right) + \frac{2}{C^2}.
\end{equation}

We find that the Kretschmann scalar for the interior space-time is given by a lengthy expression and has $\sim 30$ non-zero components \cite{Lake}. It involves inverse powers of the field functions $A$, $B$ and $C$. Therefore, both the Ricci and the Kretschmann scalars will diverge at the time $t=t_s$, when these functions tend to zero, indicating a possible curvature singularity. However, for an exact analysis we need to find the solution to the $f(R,T)$ field equations in our case, a fairly elaborate program which will be presented elsewhere.

The $f(R,T)$ field equations (\ref{EFTfull}) for the spherical non-static interior (\ref{intmetric}) are found as
\begin{equation}
\label{A}
G_{00}=\frac{A^2}{f_{R}}\left[{\rho}+{\epsilon}(1 + f_T) - \frac{1}{2}\left(f-Rf_{R}\right)+\frac{\zeta_{00}}{A^2}\right],
\end{equation}

\begin{equation}
\label{B}
G_{01}=\frac{AB}{f_{R}}\left[-(1+f_T)(q+{\epsilon}) + \frac{\zeta_{01}}{AB} \right],
\end{equation}

\begin{eqnarray}
\label{C}
G_{11}=\frac{B^2}{f_{R}}\left[(1+f_T)(p_r+\epsilon+4\eta{\sigma})+\rho f_T + \frac{1}{2}\left(f-Rf_{R}\right) + \frac{\zeta_{11}}{B^2}\right] ,
\end{eqnarray}

\begin{eqnarray}
\label{D}
G_{22}=\frac{C^2}{f_{R}}\left[(1+f_T)({p_{\perp}}-2\eta{\sigma})+\rho f_T + \frac{1}{2}\left(f-Rf_{R}\right) + \frac{\zeta_{22}}{C^2}\right],
\end{eqnarray}
where
\begin{eqnarray}
  \zeta_{00} &=& \left[ -\left(\frac{\dot{B}}{B}+\frac{2\dot{C}}{C}\right)\dot{f}_R +\frac{A^2}{B^2}\left(\left(\frac{2C'}{C}-\frac{B'}{B}\right)f'_R + f''_R \right) \right], \\
  \zeta_{01} &=& \left[\dot{f}'_R-\frac{A'}{A}\dot{f}_R-\frac{\dot{B}}{B}f'_R \right], \\
  \zeta_{11} &=& \left[\frac{B^2}{A^2}\left(\ddot{f}_R + \left(\frac{2\dot{C}}{C}-\frac{\dot{A}}{A}\right)\dot{f}_R \right) -\left(\frac{A'}{A}+\frac{2C'}{C}\right)f'_R \right], \\
  \zeta_{22} &=& \left[\frac{C^2}{A^2} \left(\ddot{f}_R + \left(-\frac{\dot{A}}{A}+\frac{\dot{B}}{B}+\frac{\dot{C}}{C}\right)\dot{f}_R \right) +\frac{C^2}{B^2}\left(\left(-\frac{A'}{A}+\frac{B'}{B}-\frac{C'}{C}\right)f'_R -f''_R \right) \right].
\end{eqnarray}

\section{Dynamical Equations for the collapsing system}
In his pioneering work, Chandrasekhar showed within the framework of GR \cite{C2}, that even under conditions of hydrostatic equilibrium, a self-gravitating gaseous system becomes dynamically unstable (with respect to radial oscillations) much before they reach the `Schwarzschild limit'. In Newtonian gravity, the condition of dynamical instability is related to the ``ratio of specific heats'' of the system, and a difference in the nature of the gravitational field --- general relativistic or other, decides the exact condition of stability of such a system. In fact, different degrees of instability will affect the nature of evolution of the collapsing object, thereby leading to a difference in the end state of collapse \cite{HS}. The problem of stability has therefore been widely investigated \cite{YBF, YBBF, YBF2}, and is done by examining the dynamical equations of collapse.

The dynamical equations describe the evolution of the parameters of the collapsing stellar object with time and radius. These equations are derived from the conservation laws. We consider the conservation of the Einstein tensor because the matter energy momentum tensor has a non-vanishing divergence in $f(R,T)$ gravity, as is evident from (\ref{divEMT}). The divergence of the effective energy-momentum tensor in (\ref{EFTfull}) provides the following continuity equations:

\begin{eqnarray}
\nonumber && \frac{\dot{f_{R}}}{f_{R}^{2}A}\left[\rho+\epsilon\left(1+f_{T}\right)-\frac{f}{2}\right]+\frac{\left(q+\epsilon\right)\left(1 +f_{T}\right)}{Bf_{R}}\left[\frac{f_{R}'}{f_{R}}-\frac{4f_{T}}{\left(-1+f_{T}\right)}\left(\frac{A'}{A}+\frac{C'}{C}\right)\right]+\frac{\dot{f} -\dot{R}f_{R}}{2Af_{R}} \\
\nonumber && +\frac{f_{T}}{f_{R}A\left(-1+f_{T}\right)}\left[-\frac{2A}{B}\left(1+f_{T}\right)\left(\epsilon'+q'\right)-2\left(\dot{\rho} +\frac{f_{T}'A\left(q+\epsilon\right)}{B}\right)+\frac{\left(1+f_{T}\right)\dot{T}}{2}-2\dot{f_{T}}\epsilon-2\left(1+f_{T}\right)\dot{\epsilon}\right. \\
\label{cont_1} && \left.-2\left(1+f_{T}\right)\left[\frac{\dot{B}}{B}\left(p_{r}+\rho+2\epsilon+4\eta\sigma\right)+\frac{\dot{C}}{C}\left(\rho+p_{\perp}+\epsilon -2\eta\sigma\right)\right]\right]+\mathscr{Z}_{1}(t,r)=0, \\
\nonumber \textrm{and} \\
\nonumber && \frac{8\eta f_{T}\left(1+f_{T}\right)\left(\sigma A\right)'}{ABf_{R}\left(1-f_{T}\right)}-\frac{4\eta\left(1+f_{T}\right)f_{R}'\sigma}{Bf_{R}^{2}}-\frac{f_{R}'}{Bf_{R}^{2}}\left[\left(p_{r}+\rho +\epsilon\right)f_{T}+p_{r}+\epsilon+\frac{f}{2}\right]-\frac{\dot{f_{R}}}{Af_{R}^{2}}\left(q+\epsilon\right)\left(1+f_{T}\right) \\
\nonumber && -\frac{4}{AB^{3}C^{2}f_{R}\left(1-f_{T}\right)}\left[f_{T}\left(1 +f_{T}\right)\left(6\eta B^{2}CC'A\sigma+\left(q+\epsilon\right)B^{3}C^{2}\left(\frac{\dot{B}}{B}+\frac{\dot{C}}{C}\right) +B^{2}C'AC\left(p_{r}-p_{\perp}+\epsilon\right)\right)\right.  \\
\nonumber && +2A\eta\sigma f_{T}f_{T}'B^{2}C^{2} + \frac{B^{2}C^{2}}{2} \left\{ A'f_{T}\left(1+f_{T}\right)\left(p_{r}+\rho+2\epsilon\right)+f_{T}f_{T}'A \left(p_{r}+\rho+\epsilon \right)\right. + \left(A\epsilon'+B\dot{\epsilon}\right)f_{T}\left(1+f_{T}\right)  \\
\nonumber && \left.\left. +\frac{A}{4}f_{T}\left(1+f_{T}\right)\left(T'+4p_{r}'\right) +Bf_{T}\left\{ \dot{f_{T}}\left(q+\epsilon\right)+\dot{q}\left(1+f_{T}\right) \right\} \right\} \right] +\frac{1}{Bf_{R}}\left[\rho'f_{T}+\frac{1}{2}\left(f'-f_{R}R'\right)\right] +\mathscr{Z}_{2}(t,r)=0,\\
\label{cont_2}
\end{eqnarray}
where the dark source terms are
\begin{eqnarray}
\nonumber  \mathscr{Z}_{1}(t,r) &=& -\frac{\dot{f_{R}}A''}{A^{2}B^{2}f_{R}}+\frac{\dot{f_{R}}\ddot{B}}{Bf_{R}A^{3}}-\frac{2f_{R}'\dot{C}'}{B^{2}f_{R}AC}+\frac{2\dot{f_{R}}\ddot{C}}{Cf_{R}A^{3}} +\frac{\dot{f_{R}}f_{R}''}{AB^{2}f_{R}^{2}}-\frac{f_{R}'\dot{f_{R}}'}{AB^{2}f_{R}^{2}} -\frac{\left(2B\dot{C}+\dot{B}C\right)\dot{f_{R}}^{2}}{A^{3}BCF^{2}}-\frac{\dot{f_{R}}\dot{A}\dot{B}}{BA^{4}f_{R}}  \\
\label{cont_1a} & &  -\frac{2\dot{A}\dot{C}\dot{f_{R}}}{A^{4}f_{R}C}+\frac{\dot{f_{R}}A'f_{R}'}{A^{2}B^{2}f_{R}^{2}}+\frac{2\dot{C}f_{R}'A'}{A^{2}B^{2}f_{R}C} +\frac{\left(B'C-2BC'\right)\left(f_{R}A'-Af_{R}'\right)\dot{f_{R}}}{A^{2}B^{3}f_{R}^{2}C}+\frac{\left(Cf_{R}' +2C'f_{R}\right)f_{R}'\dot{B}}{AB^{3}f_{R}^{2}C} , \\
\textrm{and} \nonumber \\
\nonumber  \mathscr{Z}_{2}(t,r) &=& \frac{2\dot{f_{R}}\dot{C}'}{A^{2}Bf_{R}C}-\frac{f_{R}'A''}{AB^{3}f_{R}}+\frac{f_{R}'\ddot{B}}{A^{2}B^{2}f_{R}}-\frac{2f_{R}'C''}{B^{3}f_{R}C} +\frac{\dot{f_{R}}\dot{f_{R}}'}{A^{2}Bf_{R}^{2}}-\frac{\ddot{f_{R}}f_{R}'}{A^{2}Bf_{R}^{2}} +\frac{f_{R}'^{2}\left(A'C+2AC'\right)}{AB^{3}Cf_{R}^{2}}-\frac{f_{R}'\dot{B}\left(\dot{f_{R}}A +\dot{A}f_{R}\right)}{A^{3}B^{2}f_{R}^{2}}  \\
\label{cont_2a} & & +\frac{2\dot{B}\dot{C}f_{R}'}{A^{2}B^{2}f_{R}C} -\frac{2C'\dot{B}\dot{f_{R}}}{A^{2}B^{2}f_{R}C} +\frac{2B'C'f_{R}'}{B^{4}f_{R}C}-\frac{2f_{R}'\dot{f_{R}}\dot{C}}{A^{2}BCf_{R}^{2}} +\frac{f_{R}'\dot{f_{R}}\dot{A}}{A^{3}Bf_{R}^{2}}+\frac{f_{R}'A'B'}{AB^{4}f_{R}}-\frac{A'\dot{f_{R}}^{2}}{A^{3}Bf_{R}^{2}} -\frac{2\dot{C}\dot{f_{R}}A'}{A^{3}Bf_{R}C} .
\end{eqnarray}
The parameters $\mathscr{Z}_{1}$ and $\mathscr{Z}_{2}$ are the extra curvature degrees of freedom representing the dark sources that arise automatically from the $f(R,T)$ gravitational field. These terms signify the corrections in the variation of total energy of the self-gravitating relativistic matter across its boundaries as time evolves. Rearranging the dark source terms, we can represent them in the following form:
\begin{align*}
\mathscr{Z}_{1} & =\frac{1}{A}\left[-\frac{1}{A^{2}}\left\{\left(\frac{A^{2}}{f_{R}}\left[-\frac{f-Rf_{R}}{2}+\frac{D_{00}}{A^{2}}\right]\right)_{,0} -\frac{2\dot{A}A}{f_{R}}\left(-\frac{f-Rf_{R}}{2}+\frac{D_{00}}{A^{2}}\right)-\frac{2A'D_{00}}{Bf_{R}}\right\}\right.\\
 & \left. +\frac{1}{B^{2}}\left\{\left(\frac{D_{01}}{f_{R}}\right)_{,1} -\frac{B\dot{B}}{f_{R}}\left(-\frac{f-Rf_{R}}{2}+\frac{D_{00}}{A^{2}}\right)-\left(\frac{B'}{B}+\frac{A'}{A}\right)\frac{D_{01}}{f_{R}} -\frac{B\dot{B}}{f_{R}}\left(\frac{f-Rf_{R}}{2}+\frac{D_{11}}{B^{2}}\right)\right\}\right],
\end{align*}
and
\begin{align*}
\mathscr{Z}_{2} & =\frac{1}{B}\left[-\frac{1}{A^{2}}\left\{\left(\frac{D_{01}}{f_{R}}\right)_{,0}-\left(\frac{\dot{B}}{B} +\frac{\dot{A}}{A}\right)\frac{D_{01}}{f_{R}} -\frac{AA'}{f_{R}}\left(\frac{f-Rf_{R}}{2} +\frac{D_{11}}{B^{2}}\right)-\frac{AA'}{f_{R}}\left(-\frac{f-Rf_{R}}{2}+\frac{D_{00}}{A^{2}}\right)\right\}\right.\\
 & \left. +\frac{1}{B^{2}}\left\{\left(\frac{B^{2}}{f_{R}}\left[\frac{f-Rf_{R}}{2}+\frac{D_{11}}{B^{2}}\right]\right)_{,1} -\frac{2B\dot{B}D_{01}}{A^{2}f_{R}}-\frac{2BB'}{f_{R}}\left(\frac{f-Rf_{R}}{2}+\frac{D_{11}}{B^{2}}\right)\right\}\right].
\end{align*}

The Misner-Sharp mass function \cite{MS} representing the total gravitational energy entrapped inside the surface $\Sigma$ bounding the spherical star of radius `C', is given by
\begin{eqnarray}
\label{eq:g8c} M(v)_{\Sigma} &=& \left[\frac{C}{2}\left(1 + \frac{\dot{C}^2}{A^2}-\frac{{C'}^2}{B^2}\right)\right]_{\Sigma}.
\end{eqnarray}
In order to calculate the variation of this mass function through the boundary surface of the collapsing configuration, we define two well-known operators: the proper time derivative
\begin{equation}\label{ptd}
D_{T}=\frac{1}{A}\frac{\partial}{\partial{t}},
\end{equation}
and the proper radial derivative $D_{C}$ (constructed from the radius of the sphere inside $\Sigma$)
\begin{equation}\label{prd}
D_{C}=\frac{1}{C'}\frac{\partial}{\partial{r}}.
\end{equation}
The relativistic 4-velocity of the fluid for the corresponding collapse is given by
\begin{equation}\label{rfv}
U=D_{T}C=\frac{\dot{C}}{A},
\end{equation}
which must be negative to ensure collapse to occur.

Defining new variable $H(t,r)=\frac{C'}{B}$, we obtain from (\ref{eq:g8c}) and (\ref{rfv})
\begin{equation}\label{ratio_E}
H=\left[1+U^{2}-\frac{2M}{C}\right]^{1/2}.
\end{equation}
With the help of the field equations and equations (\ref{prd})-(\ref{ratio_E}), the mass variation in the radial direction is found to be
\begin{equation}\label{rmv}
D_C M=\frac{C^2}{2f_R}\left[\rho +\epsilon(1+f_T) +\frac{1}{2}\left(f-Rf_{R}\right)+\frac{\zeta_{00}}{A^2}+\frac{U}{H}\left\{(1+f_T)(q+\epsilon)
-\frac{\zeta_{01}}{AB}\right\}\right],
\end{equation}
which on integration yields the result
\begin{equation}\label{rmv_intgrn}
M=\frac{1}{2}\int^C_{0}\frac{C^2}{f_R}\left[\rho +\epsilon(1+f_T) +\frac{1}{2}\left(f-Rf_{R}\right)+\frac{\zeta_{00}}{A^2}+\frac{U}{H} \left\{(1+f_T)(q+\epsilon)-\frac{\zeta_{01}}{AB}\right\}\right]dC.
\end{equation}
The temporal variation of the mass function is obtained as
\begin{equation}\label{tmv}
D_T M=-\frac{C^2}{2f_R}\left[U\left\{(1+f_T)(p_r+\epsilon+4\eta\sigma)+\rho f_T+\frac{1}{2}\left(f-Rf_{R}\right)+\frac{\zeta_{11}}{B^2}\right\}
+H\left\{(1+f_T)(q+\epsilon)-\frac{\zeta_{01}}{AB}\right\}\right].
\end{equation}

In order to determine the variation of the physical parameters of the self-gravitating system in course of time, we now consider a suitable perturbation scheme.

\section{$f(R,T)$ function and Perturbation Scheme}

To find the approximate solution of a differential equation, very often one uses the so-called `perturbation method'. This technique has been successfully applied to analyze the dynamics of gravitational collapse by several authors \cite{10,11,12,13,14,15,16,Santos,24,SZ1,SY,NZ1,NZ2,NZ3,SW,SS}. We use this technique to analyze the evolution of the stellar object of our study under the effect of $f(R,T)$ model of gravity. To apply the perturbation theory we assume that initially the matter is in a state of hydrostatic equilibrium, so that the initial values of the parameters depend only on the radial coordinate. But in course of time the perturbed quantities gather both radial and time dependence. The radial heat flow $q$ is of the order of $\varepsilon$ ($0<\varepsilon\ll1$) \cite{10}, (where $\epsilon$ and $\varepsilon$ are different quantities). We consider a combination of the Starobinsky model \cite{Starobinsky} and a linear term of the trace $T$ as the $f(R, T)$ model \cite{SZ2}, which can be written as
\begin{eqnarray}\label{m}
&&f(R, T)= R+\alpha R^2+\lambda T,
\end{eqnarray}
where $\alpha$ has positive real values, while $\lambda$ is a coupling parameter, and $\lambda T$ represents the extent of modification to the Starobinsky
$f(R)$ gravity. Thus the metric functions and the physical parameters may be written as follows:

\begin{eqnarray}
\label{41} A(t,r)&=&A_0(r)+\varepsilon D(t)a(r),
\\\label{42} B(t,r)&=&B_0(r)+\varepsilon D(t)b(r),
\\\label{43} C(t,r)&=&C_0(r)+\varepsilon D(t)\bar{c}(r),
\\\label{44} \rho(t,r)&=&\rho_0(r)+\varepsilon {\bar{\rho}}(t,r),
\\\label{45} p_r(t,r)&=&p_{r0}(r)+\varepsilon{\bar{p}_r}(t,r),
\\\label{46} p_\bot(t,r)&=&p_{\bot 0}(r)+\varepsilon {\bar{p}_\bot}(t,r)
\\\label{47} \epsilon(t,r)&=&\varepsilon {\bar{\epsilon}}(t,r)
\\\label{48} m(t,r)&=&m_0(r)+\varepsilon {\bar{m}}(t,r),
\\\label{49'} R(t,r)&=&R_0(r)+\varepsilon D_1(t)e_1(r),
\\\label{50'} T(t,r)&=&T_0(r)+\varepsilon D_2(t)e_2(r),
\\\nonumber f(R,T)&=&[R_0(r)+\alpha R_0^2(r)+\lambda T_0]+\varepsilon D_1(t)e_1(r)
\\\label{51'}&&\times[1+2\alpha R_0(r)]+\varepsilon\lambda D_2(t)e_2(r),
\\\label{52'}f_R&=&[1+2\alpha R_0(r)]+2\alpha\varepsilon D_1(t)e_1(r),
\\\label{53'} f_T&=&\lambda,
\\\label{54'} \Theta_1(t,r)&=&\varepsilon\bar{\Theta}_1,
\\\label{55'} \sigma(t,r)&=&\varepsilon {\bar{\sigma}}(t,r),
\\\label{56'} q(t,r)&=&\varepsilon\bar{q}(t,r).
\end{eqnarray}
For $\bar{c}(r)=0,$ we get the shear-free case.
Without loss of generality, we consider $C_0(r)=r$ as the Schwarzschild coordinate. In the above equations, $R_0$ represents the static part of the Ricci scalar which has the following form
\begin{eqnarray}
R_{0}(r)=-\frac{2}{rB_0^{2}}\left(\frac{1}{r}-\frac{2B'_0}{B_0}+\frac{2A'_0}{A_0}\right) - \frac{2}{B_0^{2}}\left(\frac{A''_0}{A_0}-\frac{A'_0 B'_0}{A_0 B_0}\right) + \frac{2}{r^2},
\end{eqnarray}
while the value of perturbed part of Ricci scalar is given by
\begin{equation}\label{pert_Ricciscalar}
De=\frac{2\ddot{D}}{A_{0}^{2}}\left(\frac{b}{B_{0}}+\frac{\bar{c}}{r}\right)+D\left[-\frac{2}{B_{0}^{2}}\left\{ \frac{A_{0}'}{A_{0}}\left(\frac{2\bar{c}'}{r}-\frac{b'}{B_{0}}\right)-\frac{B_{0}'}{B_{0}}\left(\frac{2\bar{c}'}{r}+\frac{a'}{A_{0}}\right) +\frac{2}{r}\left(\frac{a'}{A_{0}}-\frac{b'}{B_{0}}+\frac{\bar{c}}{r}\right)+\frac{a''}{A_{0}}+\frac{\bar{c}''}{r}\right\} \right],
\end{equation}
where we have assumed that the perturbed quantities $D_1=D_2=D$, and $e_1=e_2=e$.
The matter content of the spherical collapsing configuration under static conditions as well as non-static equilibrium phases, are given by
\begin{eqnarray}
  m_0 &=& \frac{r}{2}\left(1 - \frac{1}{B_0^2}\right), \\
  {\bar{m}}(t,r) &=& \frac{D(t)}{B_0^2} \left[r \left(\frac{b}{B_0} - \bar{c}' \right) + \frac{\bar{c}}{2}\left(B_0^2 - 1 \right)  \right].
\end{eqnarray}

Let $Z=1+2\alpha R_{0}$, $Z_{R_{0}}=\frac{\partial Z}{\partial R_{0}}$, and $Z_{R_{0}R_{0}}=\frac{\partial^{2}Z}{\partial R_{0}^{2}}$. Using these symbols, the static configuration of $f(R,T)$ field equations is obtained in the form
\begin{eqnarray}\label{55}
\frac{1}{r^{2}}\left(1-\frac{1}{B_{0}^{2}}\right)+\frac{2B_{0}'}{rB_{0}^{3}}&=&\frac{1}{Z}\left[\rho_{0}+\frac{\alpha R_{0}^{2}-\lambda T_{0}}{2}+\frac{D_{00}^{(S)}}{A_{0}^{2}}\right],\\\label{56}
\frac{1}{r^{2}}\left(\frac{1}{B_{0}^{2}}-1\right)+\frac{2A_{0}'}{A_{0}B_{0}^{2}r}&=&\frac{1}{Z}\left[p_{r_{0}}+\lambda\left(\rho_{0}+p_{r_{0}}\right) -\frac{\alpha R_{0}^{2}-\lambda T_{0}}{2}+\frac{D_{11}^{(S)}}{B_{0}^{2}}\right],\\\label{57}
\frac{1}{B_{0}^{2}}\left[\frac{A_{0}''}{A_{0}}+\frac{A_{0}'}{A_{0}r}-\frac{A_{0}'B_{0}'}{A_{0}B_{0}}-\frac{B_{0}'}{B_{0}r}\right] &=&\frac{1}{Z}\left[p_{\perp_{0}}+\lambda\left(\rho_{0}+p_{\perp_{0}}\right)-\frac{\alpha R_{0}^{2}-\lambda T_{0}}{2}+\frac{D_{22}^{(S)}}{r^{2}}\right],
\end{eqnarray}
where
\begin{eqnarray}\label{58}
D_{00}^{(S)}&=&\frac{A_{0}^{2}}{B_{0}^{2}}\left[Z_{R_{0}}\left\{ R_{0}''-R_{0}'\left(\frac{B_{0}'}{B_{0}}-\frac{2}{r}\right)\right\} +Z_{R_{0}R_{0}}\left(R_{0}'\right)^{2}\right],\\\label{59}
D_{11}^{(S)}&=&-Z_{R_{0}}R_{0}'\left(\frac{A_{0}'}{A_{0}}+\frac{2}{r}\right),\\\label{59'}
D_{22}^{(S)}&=&\frac{r^{2}}{B_{0}^{2}}\left[Z_{R_{0}}\left\{ R_{0}'\left(\frac{B_{0}'}{B_{0}}-\frac{A_{0}'}{A_{0}}-\frac{1}{r}\right)-R_{0}''\right\} -Z_{R_{0}R_{0}}\left(R_{0}'\right)^{2}\right].
\end{eqnarray}
In the static configuration, the first dynamical
equation is identically satisfied, while the second equation is given by
\begin{eqnarray}
\nonumber
&&-\frac{Z_{R_{0}}R_{0}'}{B_{0}Z^{2}}\left\{ p_{r_{0}}\left(1+\lambda\right)+\rho_{0}\lambda+\frac{R_{0}+\alpha R_{0}^{2}+\lambda T_{0}}{2}\right\} +\frac{4\lambda\left(1+\lambda\right)}{B_{0}Zr\left(-1+\lambda\right)}\left(p_{r_{0}}-p_{\perp_{0}}\right)
+\frac{\lambda\left(1+\lambda\right)T_{0}'}{2\left(-1+\lambda\right)B_{0}Z}\\
\label{B1s*}&&+\frac{2\lambda}{\left(-1+\lambda\right)}\frac{1}{B_{0}Z}\left\{ \frac{A_{0}'}{A_{0}}\left(p_{r_{0}}+\rho_{0}\right)\left(1+\lambda\right)+p_{r_{0}}'\left(1+\lambda\right)+\rho_{0}'\lambda\right\} +\mathscr{Z}_{2}^{(S)}=0,
\end{eqnarray}
where $\mathscr{Z}_{2}^{(S)}$ represents the static part of $\mathscr{Z}_{2}(t,r)$ and has the following form:
\begin{eqnarray}\label{B2s*}
\mathscr{Z}_{2}^{(S)}=\frac{Z_{R_{0}}R_{0}'}{B_{0}^{3}Z}\left[-\frac{A_{0}''}{A_{0}}+\frac{Z_{R_{0}}R_{0}'\left(A_{0}'r +2A_{0}\right)}{A_{0}rZ}+\frac{2B_{0}'}{rB_{0}}+\frac{A_{0}'B_{0}'}{A_{0}B_{0}}\right].
\end{eqnarray}
After the application of the perturbation scheme, the perturbed form of the first dynamical equation is
\begin{align}\label{59a}
-\frac{2\lambda\dot{\bar{\rho}}}{\left(-1+\lambda\right)A_{0}Z}+\frac{\left(\bar{q} +\bar{\epsilon}\right)\left(1+\lambda\right)}{B_{0}Z}\left[\frac{Z_{R_{0}}R_{0}'}{Z} -4\frac{\lambda}{\left(-1+\lambda\right)}\left(\frac{A_{0}'}{A_{0}}+\frac{1}{r}\right)\right] -\frac{2\lambda\left(1+\lambda\right)}{\left(-1+\lambda\right)Z}\left[\frac{\bar{\epsilon}' +\bar{q}'}{B_{0}}+\frac{\dot{\bar{\epsilon}}}{A_{0}}\right] \qquad \qquad \qquad \nonumber \\
+\dot{D}\left[\frac{2\alpha e}{Z^{2}A_{0}}\left(\rho_{0}-\frac{R_{0}+\alpha R_{0}^{2}+\lambda T_{0}}{2}\right)-\frac{2\lambda\left(1+\lambda\right)}{\left(-1+\lambda\right)ZA_{0}}\left\{ \frac{b}{B_{0}}\left(p_{r_{0}}+\rho_{0}\right)+\frac{2\bar{c}}{r}\left(\rho_{0}+p_{\perp_{0}}\right)-\frac{e}{4}\right\} +\frac{\lambda e}{2A_{0}Z}+\mathscr{Z}_{1}^{(P)}\right] & =0,
\end{align}
where $\mathscr{Z}_{1}^{(P)}$ is given in the Appendix.

The perturbed form of the second dynamical equation is
\begin{align}\label{59b}
\frac{4\eta(1+\lambda)}{B_{0}Z}\left[\frac{2\lambda}{\left(-1+\lambda\right)}\left(\bar{\sigma}'+\frac{\bar{\sigma}A_{0}'}{A_{0}} +\frac{3\bar{\sigma}}{r}\right)-\frac{\bar{\sigma} Z_{R_{0}}R_{0}'}{Z}\right] -\frac{Z_{R_{0}}R_{0}'}{B_{0}Z^{2}}\left[\left(\bar{p_{r}}+\bar{\epsilon}\right)\left(1+\lambda\right) +\lambda\bar{\rho}\right]\nonumber \\
+\frac{\lambda e'}{2B_{0}Z}-\frac{\alpha eR_{0}'}{B_{0}Z}+\frac{eZ_{R_{0}}R_{0}'}{2B_{0}Z} + \frac{2\lambda}{\left(-1+\lambda\right)} \left[\frac{2\left(1+\lambda\right)}{B_{0}Zr}\left(\bar{p_{r}}-\bar{p_{\perp}}+\bar{\epsilon}\right) +\frac{A_{0}'}{A_{0}B_{0}Z}\left(\bar{p_{r}} +\bar{\rho}+2\bar{\epsilon}\right)\right.\nonumber \\
\left.+\frac{\left(1+\lambda\right)}{A_{0}B_{0}Z}\left\{A_{0}\left(\bar{p_{r}}'+\bar{\epsilon}'\right)+B_{0}\left(\dot{\bar{\epsilon}} +\dot{\bar{q}}\right)\right\} +\frac{\bar{\rho}'\lambda}{B_{0}Z}\right] +\frac{\ddot{D}bZ_{R_{0}}R_{0}'}{A_{0}^{2}B_{0}^{2}Z}\nonumber \\
+D\left[-\frac{Z_{R_{0}}R_{0}'}{B_{0}Z^{2}}\frac{e}{2}\left(Z+\lambda\right)-\frac{2\alpha e'}{B_{0}Z^{2}}\left(\rho_{0}\lambda+p_{r_{0}}\left(1+\lambda\right)+\frac{R_{0}+\alpha R_{0}^{2}+\lambda T_{0}}{2}\right)\right.\nonumber \\
+\frac{2\lambda}{\left(-1+\lambda\right)}\left[\frac{2\bar{c}'}{B_{0}Zr}\left(1+\lambda\right)\left(p_{r_{0}}-p_{\perp_{0}}\right) +\frac{\left(1+\lambda\right)}{A_{0}B_{0}Z}\left[a'\left(p_{r_{0}}+\rho_{0}\right)+ap_{r_{0}}'\right]\right.\nonumber \\
\left.+\frac{a}{A_{0}B_{0}Z}\left\{ \rho_{0}'\lambda-\frac{ZR_{0}'}{2}-\frac{\left(R_{0}'+2\alpha R_{0}R_{0}'+\lambda T_{0}'\right)}{2}\right\} \right] \left.+\frac{e'\lambda\left(1+\lambda\right)}{2B_{0}Z\left(-1+\lambda\right)}+\mathscr{Z}_{2}^{(P)}\right]=0,
\end{align}
where $\mathscr{Z}_{2}^{(P)}$ is given in the Appendix. From $(\ref{59a})$, integrating with respect to $t$, we obtain,
\begin{align}\label{60}
\bar{\rho} & =\frac{\left(-1+\lambda\right)}{2\lambda}\frac{A_{0}}{B_{0}}\left[\frac{Z_{R_{0}}R_{0}'}{Z} -\frac{4\lambda}{\left(-1+\lambda\right)}\left(\frac{A_{0}'}{A_{0}} +\frac{1}{r}\right)\right]\left(1+\lambda\right)\int\left(\bar{q}+\bar{\epsilon}\right)dt -\left(1+\lambda\right)\frac{A_{0}}{B_{0}}\int\left(\bar{\epsilon}'+\bar{q}'\right)dt-\left(1+\lambda\right)\bar{\epsilon}\nonumber \\
 & +D\frac{\left(-1+\lambda\right)}{2\lambda}\left[\frac{2\alpha e}{Z}\left(\rho_{0}-\frac{R_{0}+\alpha R_{0}^{2}+\lambda T_{0}}{2}\right)-\frac{2\lambda\left(1+\lambda\right)}{\left(-1+\lambda\right)}\left\{ \frac{b}{B_{0}}\left(p_{r_{0}}+\rho_{0}\right)+\frac{2\bar{c}}{r}\left(\rho_{0}+p_{\perp_{0}}\right)-\frac{e}{4}\right\} +\frac{\lambda e}{2}+A_{0}Z\mathscr{Z}_{1}^{(P)}\right].
\end{align}
The perturbed quantities $\bar{\rho}$ and $\bar{p}_r$ are related through the ratio of specific heats,
if we consider the second law of thermodynamics and assume a Harrison-Wheeler type equation of state \cite{HW}, expressed in the following form:
\begin{eqnarray}\label{60a}
\bar{p}_r&=&\frac{\Gamma p_{r0}}{\rho_{0}+p_{r_{0}}}\bar{\rho},
\end{eqnarray}
where $\Gamma$ is the adiabatic index which determines the change of pressure for a given change in density. We assume that $\Gamma$ remains constant while the collapse progresses. Now substituting $\bar{\rho}$ from \eqref{60} in \eqref{60a}, we obtain
\begin{align}\nonumber
\bar{p}_{r} & = \frac{\Gamma p_{r_{0}}}{\rho_{0}+p_{r_{0}}}\left[\frac{\left(-1+\lambda\right)}{2\lambda}\frac{A_{0}}{B_{0}}\left[\frac{Z_{R_{0}}R_{0}'}{Z} -\frac{4\lambda}{\left(-1+\lambda\right)}\left(\frac{A_{0}'}{A_{0}}+\frac{1}{r}\right)\right]\left(1+\lambda\right)\int\left(\bar{q} +\bar{\epsilon}\right)dt-\frac{A_{0}}{B_{0}}\left(1+\lambda\right)\int\left(\bar{\epsilon}'+\bar{q}'\right)dt -\left(1+\lambda\right)\bar{\epsilon}\right.\nonumber \\
& \left.+D\frac{\left(-1+\lambda\right)}{2\lambda}\left[\frac{2\alpha e}{Z}\left(\rho_{0}-\frac{R_{0}+\alpha R_{0}^{2}+\lambda T_{0}}{2}\right)-\frac{2\lambda\left(1+\lambda\right)}{\left(-1+\lambda\right)}\left\{ \frac{b}{B_{0}}\left(p_{r_{0}}+\rho_{0}\right)+\frac{2\bar{c}}{r}\left(\rho_{0}+p_{\perp_{0}}\right)-\frac{e}{4}\right\} +\frac{\lambda e}{2}+A_{0}Z\mathscr{Z}_{1}^{(P)}\right]\right]. \label{61}
\end{align}
Considering the perturbed variant of the field equation \eqref{D}, we obtain the following form of the perturbed term $\bar{p}_{\perp}$:
\begin{equation}\label{62}
\bar{p}_{\perp}=-\frac{\ddot{D}Z\left(\bar{c}+\frac{br}{B_{0}}\right)}{A_{0}^{2}r\left(1+\lambda\right)}-\frac{\bar{\rho}\lambda}{1+\lambda}+2\eta\bar{\sigma} +DH_{\perp},
\end{equation}
where $H_{\perp}$ is given in the Appendix. Substituting \eqref{60}, \eqref{61}, and \eqref{62}, in \eqref{59b}, we get an equation of the form
\begin{equation}\label{63}
\ddot{D}-Q(r)D=G(r,t),
\end{equation}
where $Q(r)$ and $G(r,t)$ are given in the Appendix. The terms in \eqref{63} are chosen
to ensure that $Q(r)$ remains positive throughout the evolution. The solution of this differential equation is given by
\begin{equation}\label{64}
D=\frac{1}{2\sqrt{Q}}\left[e^{\sqrt{Q}t}\int Ge^{-\sqrt{Q}t}dt-e^{-\sqrt{Q}t}\int Ge^{\sqrt{Q}t}dt\right].
\end{equation}
The explicit form of equation \eqref{63} is given in equation \eqref{explicit} in the appendix.

\subsection{Conditions of instability}

We now examine the instability conditions of the collapsing system by analysing the Newtonian and post Newtonian regimes of the collapse equations. This analysis also reveals the significance of the adiabatic index $\Gamma$ in the collapse dynamics.

\subsubsection{Newtonian Approximation}

To deal with the Newtonian Approximation of the collapsing configuration, let us consider $A_{0}=1$, $B_{0}=1$. Along with this we assume that
$\rho_{0}\gg p_{r_{0}}$, and $\rho_{0}\gg p_{\perp_{0}}$. Putting
these assumptions in the explicit form \eqref{explicit} of equation \eqref{63}, we find that the range of stability of the collapsing matter is specified by the following inequality:
\begin{equation}\label{65}
\Gamma<\frac{ZH_{1}\left(-1+\lambda\right)}{2\lambda p_{r_{0}}\left(\frac{\bar{\rho}_{N}}{\rho_{0}}\right)'\left(1+\lambda\right)},
\end{equation}
where $H_1$ is given in the appendix. The collapse will become unstable unless this condition is satisfied. We then have
\begin{align}\label{66}
\bar{\rho}_{N} & =\frac{\left(-1+\lambda\right)}{2\lambda}\left[\frac{Z_{R_{0}}R_{0}'}{Z} -\frac{4\lambda}{\left(-1+\lambda\right)}\frac{1}{r}\right]\left(1+\lambda\right)\int\left(\bar{q}+\bar{\epsilon}\right)dt -\left(1+\lambda\right)\int\left(\bar{\epsilon}'+\bar{q}'\right)dt-\left(1+\lambda\right)\bar{\epsilon}\nonumber \\
 & +D\frac{\left(-1+\lambda\right)}{2\lambda}\left[\frac{2\alpha e}{Z}\left(\rho_{0}-\frac{R_{0}+\alpha R_{0}^{2}+\lambda T_{0}}{2}\right)-\frac{2\lambda\left(1+\lambda\right)}{\left(-1+\lambda\right)}\left[\rho_{0}\left(b+\frac{2\bar{c}}{r}\right) -\frac{e}{4}\right]+\frac{\lambda e}{2}+Z\mathscr{Z}_{1_{N}}^{(P)}\right].
\end{align}
To ensure the positivity on the right hand side of the inequality \eqref{65}, we need the following constraints:
\begin{align}\label{69}
 \frac{2\lambda}{\left(-1+\lambda\right)}\left(\bar{\sigma}'+\frac{2\bar{\sigma}}{r}\right)<\frac{\bar{\sigma}Z_{R_{0}}R_{0}'}{Z}, \\  \bar{\epsilon}\left(1+\lambda\right)+\lambda\bar{\rho}_{N}>0, \\  \frac{Z_{R_{0}}R_{0}'}{Z^{2}}>\frac{2\lambda}{\left(-1+\lambda\right)}\frac{2}{Zr}, \\
e^{\sqrt{Q}t}\int Ge^{-\sqrt{Q}t}dt>e^{-\sqrt{Q}t}\int Ge^{\sqrt{Q}t}dt, \\
G+\frac{\sqrt{Q}}{2}\left[e^{\sqrt{Q}t}\int Ge^{-\sqrt{Q}t}dt-e^{-\sqrt{Q}t}\int Ge^{\sqrt{Q}t}dt\right]  <0,\\
\frac{2\left(\bar{c}+br\right)}{r^{2}}+\frac{\left(-1+\lambda\right)}{2\lambda}\frac{bZ_{R_{0}}R_{0}'}{Z}>0,\\
0<p_{r_{0}}<p_{\perp_{0}},\\
\bar{\epsilon}'+\dot{\bar{\epsilon}}+\dot{\bar{q}}<0,\\
\frac{\lambda\bar{\rho}_{N}'}{Z}<0,\\
\lambda\rho_{0}+\frac{R_{0}+\alpha R_{0}^{2}+\lambda T_{0}}{2}>0,\\
\frac{Z_{R_{0}}R_{0}'e\left(Z+\lambda\right)}{2Z^{2}}>0,\\
\frac{a'\left(1+\lambda\right)\rho_{0}}{Z}+\frac{\lambda e'}{2Z}-\frac{\alpha eR_{0}'}{Z}+\frac{eZ_{R_{0}}R_{0}'}{2Z}<0,\\
\rho_{0}'\lambda-\frac{ZR_{0}'}{2}-\frac{\left(R_{0}'+2\alpha R_{0}R_{0}'+\lambda T_{0}'\right)}{2}<0,\\
\mathscr{Z}_{2_{N}}^{(P)}<0,\\
\left(\frac{\bar{\rho}_{N}}{\rho_{0}}\right)'>0.
\end{align}

\subsubsection{Post-Newtonian Approximation}

For the Post-Newtonian Approximation, we consider $A_{0}=1-\frac{m_{0}}{r}$,
and $B_{0}=1+\frac{m_{0}}{r}$. So, we have
$$\frac{A_{0}}{B_{0}}=\frac{r-m_{0}}{r+m_{0}}, \qquad \frac{A_{0}'}{A_{0}}=\frac{m_{0}}{r\left(r-m_{0}\right)}, \qquad \textrm{and} \qquad
\frac{B_{0}'}{B_{0}}=-\frac{m_{0}}{r\left(r+m_{0}\right)}.$$
Applying these conditions to equation \eqref{explicit}, we arrive at the stability criterion
\begin{equation}\label{70}
\Gamma<\frac{\left(-1+\lambda\right)Z\left(r+m_{0}\right)\left(\rho_{0}+p_{r_{0}}\right)H_{2}}{2\lambda p_{r_{0}}\left\{ \left(1+\lambda\right)r\left(\bar{\rho}_{pN}' - \frac{\bar{\rho}_{pN}\rho_{0}'}{\left(\rho_{0} + p_{r_{0}}\right)}\right) -\bar{\rho}_{pN}\left(\frac{\left(-1+\lambda\right)}{2\lambda}\frac{Z_{R_{0}}R_{0}'r\left(1+\lambda\right)}{Z}-2\left(1+\lambda\right) -\frac{m_{0}}{\left(r-m_{0}\right)}\right)\right\} },
\end{equation}
where $H_2$ is given in the appendix. We also have
\begin{align}\label{71}
\bar{\rho}_{pN} & =\frac{\left(-1+\lambda\right)}{2\lambda}\frac{\left(r-m_{0}\right)}{\left(r+m_{0}\right)}\left[\frac{Z_{R_{0}}R_{0}'}{Z} -\frac{4\lambda}{\left(-1+\lambda\right)}\frac{1}{\left(r-m_{0}\right)}\right]\left(1+\lambda\right)\int\left(\bar{q}+\bar{\epsilon}\right)dt \nonumber \\
 & -\frac{\left(r-m_{0}\right)}{\left(r+m_{0}\right)}\left(1+\lambda\right)\int\left(\bar{\epsilon}'+\bar{q}'\right)dt
 -(1+\lambda)\bar{\epsilon} + \frac{D \left(\lambda-1 \right)}{2\lambda} \times  \left[\frac{2\alpha e}{Z}\left(\rho_{0}-\frac{R_{0}+\alpha R_{0}^{2}+\lambda T_{0}}{2}\right)\right. \nonumber \\
 & \left. \qquad \qquad -\frac{2\lambda\left(1+\lambda\right)}{\left(-1+\lambda\right)}\left\{ \frac{br}{r+m_{0}}\left(p_{r_{0}}+\rho_{0}\right)+\frac{2\bar{c}}{r}\left(\rho_{0}+p_{\perp_{0}}\right)-\frac{e}{4}\right\} +\frac{\lambda e}{2}+\left(1-\frac{m_{0}}{r}\right)Z\mathscr{Z}_{1_{pN}}^{(P)}\right].
\end{align}
To ensure that the right hand side of the inequality \eqref{70} remains positive, we impose at the following constraints:
\begin{align}\label{72}
\left(\bar{\rho}_{pN}'-\frac{\bar{\rho}_{pN}\rho_{0}'}{\left(\rho_{0} +p_{r_{0}}\right)}\right) > \bar{\rho}_{pN}\left(\frac{\left(-1+\lambda\right)}{2\lambda}\frac{Z_{R_{0}}R_{0}'}{Z}-\frac{2}{r} -\frac{m_{0}}{r\left(r-m_{0}\right)\left(1+\lambda\right)}\right),\\
r+m_{0}>0,\\
\rho_{0}+p_{r_{0}}>0,\\
\frac{2\lambda}{\left(-1+\lambda\right)}\left(\bar{\sigma}'+\frac{m_{0}\bar{\sigma}}{r\left(r-m_{0}\right)} +\frac{2\bar{\sigma}}{r}\right)<\frac{\bar{\sigma}Z_{R_{0}}R_{0}'}{Z},\\
\bar{\epsilon}\left(1+\lambda\right)+\lambda\bar{\rho}_{pN}>0,\\
\frac{Z_{R_{0}}R_{0}'r}{Z^{2}}>\frac{2\lambda}{\left(-1+\lambda\right)}\frac{2}{Z},\\
e^{\sqrt{Q}t}\int Ge^{-\sqrt{Q}t}dt>e^{-\sqrt{Q}t}\int Ge^{\sqrt{Q}t}dt,\\
G+\frac{\sqrt{Q}}{2}\left[e^{\sqrt{Q}t}\int Ge^{-\sqrt{Q}t}dt-e^{-\sqrt{Q}t}\int Ge^{\sqrt{Q}t}dt\right]  <0,\\
\frac{2r\left(\bar{c}+\frac{br^{2}}{r+m_{0}}\right)}{\left(r+m_{0}\right)\left(r-m_{0}\right)^{2}} +\frac{\left(-1+\lambda\right)}{2\lambda}\frac{bZ_{R_{0}}R_{0}'r^{4}}{\left(r^{2}-m_{0}^{2}\right)^{2}Z}>0,\\
\frac{\bar{\epsilon}'}{\left(r+m_{0}\right)}+\frac{\dot{\bar{\epsilon}}+\dot{\bar{q}}}{\left(r-m_{0}\right)}<0,\\
\frac{m_{0}}{\left(r^{2}-m_{0}^{2}\right)Z}\left(2\bar{\epsilon}+\bar{\rho}_{pN}\right)<0,\\
\frac{\lambda r\bar{\rho}_{pN}'}{\left(r+m_{0}\right)Z}<0,\\
\frac{Z_{R_{0}}R_{0}'re\left(Z+\lambda\right)}{2Z^{2}\left(r+m_{0}\right)}>0,\\
\lambda\rho_{0}+p_{r_{0}}\left(1+\lambda\right)+\frac{R_{0}+\alpha R_{0}^{2}+\lambda T_{0}}{2}>0,\\
0<p_{r_{0}}<p_{\perp_{0}},\\
\frac{a'r^{2}\left(1+\lambda\right)\left(p_{r_{0}}+\rho_{0}\right)}{\left(r^{2}-m_{0}^{2}\right)Z}+\frac{\lambda e'r}{2Z\left(r+m_{0}\right)}-\frac{\alpha erR_{0}'}{Z\left(r+m_{0}\right)}+\frac{erZ_{R_{0}}R_{0}'}{2\left(r+m_{0}\right)Z}<0,\\
\rho_{0}'\lambda-\frac{ZR_{0}'}{2}-\frac{\left(R_{0}'+2\alpha R_{0}R_{0}'+\lambda T_{0}'\right)}{2}<0,\\
\frac{e'\lambda\left(1+\lambda\right)r}{2\left(-1+\lambda\right)\left(r+m_{0}\right)Z}<0,\\
\mathscr{Z}_{2_{pN}}^{(P)}<0.
\end{align}

\section{Transport Equations}
To derive the transport equation for an arbitrary fluid, we have to
keep in mind that both the energy-momentum tensor $T^{ab}$ and the total particle flux $N^{a}$ are conserved, and the second law of
thermodynamics is obeyed by the entropy flux $S^{a}$. Mathematically,
we have
\begin{eqnarray*}
T_{;b}^{ab}=0, \qquad N_{;a}^{a}=0, \qquad S_{;a}^{a}\geq 0.
\end{eqnarray*}
There exists a unique timelike eigenvector $u_{E}^{a}$ to $T^{ab}$,
and another timelike vector $u_{N}^{a}$ parallel to $N^{a}$. In
equilibrium, a rest-frame can be defined in which these eigenvectors,
and $S^{a}$ are parallel to each other. However they are not so when the system deviates from the equilibrium state \cite{Israel1}.

Utilising the above three requirements for the energy-momentum tensor,
the particle flux and the entropy flux, the transport equation is
obtained as
\begin{equation}
\tau h^{ab}u^{c}q_{b;c}+q^{a}=-Kh^{ab}(T_{1,b}+T_{1}a_{b})-\frac{1}{2}KT_{1}^{2}\left(\frac{\tau u^{b}}{KT_{1}^{2}}\right)_{;b}q^{a}.
\end{equation}
where, $\tau$ is the relaxation time (within which the system reverts
back to the equilibrium state after being disturbed), $K$ is the
thermal conductivity, $T_{1}$ is the temperature, $h^{ab}$ is the
projection tensor, $a_{b}$ is the acceleration vector, $u^{a}$ is
the four-velocity and $q^{a}$ is the heat-flux vector. This is the causal heat transport equation derived from the M\"{u}ller-Israel-Stewart theory \cite{Muller, Israel1, Israel2, Israel3} for dissipative fluids, and used in \cite{12,PHDMS}.
Using the specified definition of the four-velocity in the comoving
frame, and radial heat flow, the transport equation can be written as
\begin{equation}
\dot{q}=-\frac{Aq}{\tau}-\frac{K(T_{1}A)'}{\tau B}-\frac{qA\Theta_{1}}{2}-\frac{qKT_{1}^{2}}{2\tau}\dot{\left(\frac{\tau}{KT_{1}^{2}}\right)}.
\end{equation}
Rearranging this equation we get
\begin{equation}
\tau\dot{q}=-\frac{qKT_{1}^{2}}{2}\dot{\left(\frac{\tau}{KT_{1}^{2}}\right)} - \tau q\left(\frac{\dot{B}}{2B}+\frac{\dot{C}}{C}\right)-\frac{K(T_{1}A)'}{B}-qA.
\end{equation}
which is equation (47) of \cite{PHDMS}. This equation can be further simplified
as
\begin{equation}
\dot{q}=-\frac{K(T_{1}A)'}{\tau B}-qA\left[\frac{1}{\tau}+\frac{\Theta_{1}}{2}+\frac{1}{2}D_{T}\left[\ln\left(\frac{\tau}{KT_{1}^{2}}\right)\right]\right],
\end{equation}
where $D_{T}=\frac{1}{A}\frac{\partial}{\partial t}$. Finally we can write the transport equation in the form
\begin{equation}\label{transport}
D_{T}q=-\frac{K(T_{1}A)'}{A\tau B}-q\left[\frac{1}{\tau}+\frac{\Theta_{1}}{2}+\frac{1}{2}D_{T}\left[\ln\left(\frac{\tau}{KT_{1}^{2}}\right)\right]\right].
\end{equation}
The collapse
velocity is given by $U=D_{T}C=\frac{\dot{C}}{A}$, which must be negative
in order to ensure collapse. Considering $D_{T}U$ as the collapse
acceleration, and the expression for the Misner-Sharp mass-energy
$M$ contained in a sphere of radius $C$, we obtain from our field
equation \eqref{C},
\begin{equation}
D_{T}U=-\frac{M}{C^{2}}-\frac{C}{2f_{R}}\left[\left(1+f_{T}\right)\left(p_{r}+\epsilon+4\eta\sigma\right)+\rho f_{T}+\frac{1}{2}\left(f-RF\right)+\frac{D_{11}}{B^{2}}\right]+\frac{A'H}{AB},
\end{equation}
where $H$ is given in \eqref{ratio_E}.

To understand the effects of dissipation on the collapse dynamics, we now combine the transport equation with the dynamical equation, remembering that, $D_{C}=\frac{1}{C'}\frac{\partial}{\partial r}$. Substituting for $\frac{A'}{A}$ and coupling $D_{T}q$ with the 2nd dynamical equation \eqref{cont_2}, and utilising all the above relations, we
get
\begin{align}\label{CoupEq}
\nonumber & \left(p_{r}+\rho+2\epsilon-4\eta\sigma\right)\left(1-\Lambda\right)D_{T}U = \left(1-\Lambda\right)F_{grav}+F_{hyd} \\
\nonumber & -H\left(q+\epsilon\right)\left[\frac{\left(1-f_{T}\right)}{2f_{R}f_{T}}D_{T}f_{R}+\frac{D_{T}f_{T}}{1+f_{T}}+2\frac{D_{T}B}{B}\right] -\frac{H^{2}\left(D_{C}f_{R}\right)\left(1-f_{T}\right)}{f_{R}f_{T}\left(1+f_{T}\right)}\left[\left(1+f_{T}\right)\left(p_{r}+\epsilon+4\eta\sigma\right)+\rho f_{T}+\frac{f}{2}\right]\\
\nonumber & -2H\left(q+\epsilon\right)\left[\frac{U}{C}\right]+\frac{H^{2}KD_{C}T_{1}}{\tau}+Hq\left[\frac{1}{\tau}+\frac{\Theta_{1}}{2}\right]\\
\nonumber & +H\left[\frac{q}{2}D_{T}\left[\ln\left(\frac{\tau}{KT_{1}^{2}}\right)\right]-D_{T}\epsilon\right] +\frac{H^{2}f_{R}\left(1-f_{T}\right)}{2f_{T}\left(1+f_{T}\right)}\left[\frac{f_{T}D_{C}\rho}{f_{R}}+\frac{D_{C}f}{2f_{R}}-D_{C}R +\frac{\mathscr{Z}_{2}\left(t,r\right)}{H}\right]\\
\end{align}
where
\begin{equation}\label{u1}
\Lambda=\frac{KT_{1}}{\tau}\left(p_{r}+\rho+2\epsilon-4\eta\sigma\right)^{-1},
\end{equation}
\begin{equation}\label{u2}
F_{grav}=-\left(p_{r}+\rho+2\epsilon-4\eta\sigma\right)\left[\frac{M}{C^{2}}+\frac{C}{2f_{R}}\left[\left(1+f_{T}\right)\left(p_{r}+\epsilon+4\eta\sigma\right)+\rho f_{T}+\frac{1}{2}\left(f-RF\right)+\frac{D_{11}}{B^{2}}\right]\right],
\end{equation}
and
\begin{equation}\label{u3}
F_{hyd}=-H^{2}\left[D_{C}\left(p_{r}+\epsilon-4\eta\sigma+\frac{T}{4}\right)+\frac{2}{C}\left(p_{r}-p_{\perp}+\epsilon+6\eta\sigma\right)\right].
\end{equation}
A close inspection of (\ref{CoupEq}) reveals that the effective inertial
mass on the left, and the passive gravitational mass on the right (which is associated with $F_{grav}$), are both reduced by a factor
of $\Lambda$, which arises as a result of heat conduction. Hence
the ratio of the gravitational mass and the inertial mass remains
unchanged, which is precisely what the weak equivalence principle
is. This result is in agreement with \cite{HPFT2009}. The hydrodynamic force term $F_{hyd}$ remains unaffected
by the thermal dissipative effects, as its coefficient is unity. It represents the gradient of the effective pressure which involves radiation pressure and also the shear viscosity. It also has a term involving the pressure anistropy, radiation and shear.
Since $U$ must be negative for collapse to occur, the term $-2H\left(q+\epsilon\right)\left[\frac{U}{C}\right]$
has a positive contribution, and denotes the matter energy which is
escaping in the form of heat-flux and radiation in the free-streaming
approximation, which reduces the collapse rate. The terms involving the temperature gradient, the expansion scalar, the thermal conductivity and the relaxation time all arise from the $D_{T}q$ term coming from the transport equation. For $\Lambda=1$, the left side of \eqref{CoupEq} vanishes, and the coefficient of $F_{grav}$ reduces to zero. Hence the passive gravitational mass density will no longer affect the collapse. Also, for values of $\Lambda>1$, the inertial mass density on the left hand side becomes negative, and the sign of $F_{grav}$ also becomes negative. Hence at the critical value of $\Lambda=1$, we expect a bounce in the system: values of $\Lambda$ less than 1 indicate increase in the inertial mass, which implies collapse of the system, while those greater than 1 implies that the system is undergoing expansion, due to a decrease in the inertial mass \cite{AA3}. The terms $\frac{H^{2}f_{R}\left(1-f_{T}\right)}{2f_{T}\left(1+f_{T}\right)}\left[\frac{D_{C}f}{2f_{R}}-D_{C}R+\frac{\mathscr{Z}_{2}\left(t,r\right)}{H}\right]$ on the right-hand side of equation (\ref{CoupEq}) arise purely from the geometry of the space-time metric and represent the effective dark energy contribution of $f(R,T)$ gravity to the collapsing process.

\section{Summary and Outlook}
In this paper, we have considered the most general non-rotating spherically symmetric metric for the interior of the collapsing matter cloud,
and have studied the dynamical instability of the dissipative collapse in the $f(R,T)$ theory of gravity. An effective measure of the instability is available from an examination of the limit for the ratio of specific heat, both in the Newtonian approximation
as well as the post-Newtonian approximation. Further, we have coupled the dynamical equations with the transport equation in order to examine the effects of dissipation on the evolution of the collapsing system.

\bigskip

Our findings from the analysis of dynamical instability are listed below:
\begin{itemize}
\item We see that in case of a non-dissipative, non-radiative fluid with
no shear, the differential equation describing the collapse becomes
a homogeneous equation and the solution is a negative exponential
in $t$ similar to the result obtained in \cite{10}. The entire expression
for $G(r,t)$ in \eqref{63} vanishes in that case.
\item In the Newtonian regime, the term $H_{1}$ will be greater for a shear-free fluid, similar to the result in
\cite{16}. It also increases if the heat flux and radiation coefficients
are zero. This is because the term involving time derivatives of the
heat flux and radiation coefficients contribute negatively to the
expression of $H_{1}$. It further increases in the case of isotropic pressure.
\item In the case of Post-Newtonian Approximation, we find that the instability
range is greater than that in the Newtonian regime, since the denominator on the right hand side of the inequality involving the ratio of specific heats decreases, and the numerator of the inequality increases. This result is also similar to
that obtained in \cite{10}. This conclusion matches with
that of Chandrasekhar \cite{C2} where it was shown that relativistic
effects increase the instability limit for the specific heat ratio.
\item Further, in the post-Newtonian regime, we also see that the
term $H_{2}$ increases if all the following quantities vanishes, viz. the shear, radiation, heat flux, and isotropic pressure.
\item However it must be mentioned that in order to ensure the positivity of
all terms in the expressions for $H_{1}$ and $H_{2}$, we have assumed that $p_{r_{0}}<p_{\perp_{0}}$, which indeed increases
the instability. However the reverse condition, i.e., $p_{r_{0}}>p_{\perp_{0}}$
will lead to a reduction in the instability compared to the case $p_{r_{0}}<p_{\perp_{0}}$
\cite{HS}.
\item We also see, from the expression for the total energy trapped inside
the fluid surface, that the heat flux and the radiation coefficient
contribute negatively towards the total energy, since their coefficient
involves $\dot{C}$ which must be negative to ensure collapse of the
fluid \cite{HS,14}. This is consistent with the result that both $H_{1}$ and $H_{2}$
would have increased if the heat flux and radiation coefficients were
put to zero.
\end{itemize}
Hence it can be concluded that the simultaneous presence of heat flux, radiation
and shear decreases the instability range for the adiabatic index, in both Newtonian and relativistic regimes. In other words,
the adiabatic index becomes more constrained in presence of dissipative, radiative and viscous effects. Further, the pressure anisotropy also
reduces the instability bound. Dynamical instability for a spherical matter distribution with pressure anisotropy, but without any shear, radiation or heat flux was studied in $f(R,T)$ theory \cite{NZ1}. These authors also studied dynamical analysis for collapse of bodies with axial symmetry and with pressure anisotropy in \cite{NZ2}, and the effect of the shear-free condition on the adiabatic index in \cite{NZ3} for the same type of matter distribution with anisotropic pressure. Although the dynamics of non-adiabatic charged spherical gravitational collapse of an anisotropic fluid with heat flux have been studied recently in the framework of $f(R,T)$ gravity \cite{AA3}, but the authors did not consider the effect of shear and free-streaming radiation, nor analysed the instability conditions. For the first time, we have studied the dynamical instability for the collapse of a more generalised fluid involving shear, radiation and heat flux and in $f(R,T)$ gravity, considering a linear coupling of the trace of the energy-momentum tensor with the Starobinsky model of $f(R)$, and showed that the instability range for the adiabatic index becomes more restricted in presence of heat flux, shear viscosity and radiation.

\bigskip

We have also derived the transport equation for the given matter in the free streaming approximation. By coupling it to the dynamical equation, we have arrived at the following results:
\begin{itemize}
  \item Both the active inertial mass and the passive gravitational mass get reduced by the same factor, in agreement with the equivalence principle, and with the works by \cite{12, HPFT2009, AbbasNazar2018}.
  \item The hydrodynamic force term remains unaffected by the thermal dissipation terms. Also, there is a reduction of mass-energy by the outgoing heat-flux and radiation, which slows down the collapse, as is evident from the negative sign before the term $-2H\left(q+\epsilon\right)\left[\frac{U}{C}\right]$, and the fact that the collapse velocity $U$ must be negative.
  \item Depending upon whether the value of $\Lambda$ in equation (\ref{CoupEq}) is greater than 1 or less than 1, the active inertial mass gets reduced or increased accordingly, and the system undergoes expansion or collapse. The value $\Lambda=1$ represents the critical point where the system undergoes a bounce between expansion and collapse. At this point, the passive gravitational mass does not affect the collapsing process \cite{AA3}. The hydrodynamic force term does not get influenced by changing the values of $\Lambda$ .
\end{itemize}
There is scope for further investigation of dynamical instability and end results of collapse of the same type of matter distribution in presence of an electromagnetic field, which will be presented in a future work. We are also investigating the thermodynamic behavior of such type of distribution.

\section*{Acknowledgments}
The initial part of this work was done earlier in IUCAA, India under the associateship programme. SG gratefully acknowledges
the warm hospitality and the facilities of work at IUCAA. We also thank Dr. Subenoy Chakraborty and Dr. Narayan Banerjee for their suggestions and comments. We thank the anonymous reviewer for the useful suggestions.

\section*{Appendix}

\begin{eqnarray}\label{73}
  \mathscr{Z}_{1}^{(P)} &=& \frac{1}{A_{0}B_{0}^{2}Z}\left[-\frac{2\alpha eA_{0}''}{A_{0}}-\frac{2\bar{c}'Z_{R_{0}}R_{0}'}{r}+\frac{2\alpha e\left\{ Z_{R_{0}R_{0}}\left(R_{0}'\right)^{2}+Z_{R_{0}}R_{0}''\right\} }{Z}-\frac{2\alpha e'Z_{R_{0}}R_{0}'}{Z}+\frac{2\alpha eA_{0}'Z_{R_{0}}R_{0}'}{A_{0}Z}\right]\nonumber \\
  & & +\frac{1}{A_{0}B_{0}^{2}Z}\left[\frac{2\alpha e\left(B_{0}'r-2B_{0}\right)\left(ZA_{0}'-A_{0}Z_{R_{0}}R_{0}'\right)}{A_{0}B_{0}Zr}+\frac{2\bar{c}A_{0}'Z_{R_{0}}R_{0}'}{A_{0}r} +\frac{bZ_{R_{0}}R_{0}'\left[rZ_{R_{0}}R_{0}'+2Z\right]}{B_{0}Zr}\right], \\
  \mathscr{Z}_{2}^{(P)} &=& \frac{2\alpha e'A_{0}''+a''Z_{R_{0}}R_{0}'}{A_{0}B_{0}^{3}Z}-\frac{2\bar{c}''Z_{R_{0}}R_{0}'}{B_{0}^{3}Zr} +\frac{\left(Z_{R_{0}}R_{0}'\right)^{2}}{A_{0}B_{0}^{3}Z^{2}r}\left(a'r+\bar{c}'A_{0}'+\bar{c}'A_{0}+2a\right) +\frac{2\alpha e'A_{0}'B_{0}'}{A_{0}B_{0}^{4}Z}\nonumber \\
  & & +\frac{4\alpha e'Z_{R_{0}}R_{0}'}{A_{0}B_{0}^{3}Z^{2}r}\left(A_{0}'r+2A_{0}\right)+\frac{2\left(b'+\bar{c}'B_{0}'\right)Z_{R_{0}}R_{0}'}{B_{0}^{4}Zr}+\frac{4\alpha e'B_{0}'}{B_{0}^{4}Zr}+\frac{Z_{R_{0}}R_{0}'\left(a'B_{0}'+b'A_{0}'\right)}{A_{0}B_{0}^{4}Z}.
\end{eqnarray}

\begin{align}\label{74}
H_{\perp} & =\frac{1}{1+\lambda}\left[\frac{Z}{r^{2}}\left(\frac{\left(a''r^{2}+2A_{0}''\bar{c}r+a'r+\bar{c}'A_{0}'r+\bar{c}A_{0}'\right)}{A_{0}B_{0}^{2}} +\frac{\bar{c}''r}{B_{0}}\right.\right. -\frac{\left(a'B_{0}'r^{2}+b'A_{0}'r^{2}+2\bar{c}rA_{0}'B_{0}'\right)}{A_{0}B_{0}^{3}} \nonumber \\
 & \left. -\frac{\left(b'r+\bar{c}'B_{0}'r+\bar{c}B_{0}'\right)}{B_{0}^{3}} -\frac{2\bar{c}r}{Z} \left\{ \left(1+\lambda\right)p_{\perp_{0}}+\rho_{0}\lambda+\frac{R_{0}+\alpha R_{0}^{2}+\lambda T_{0}-R_{0}Z}{2}+\frac{D_{22}^{(P)}}{r^{2}}\right\} \right)  \left.-\frac{e\lambda}{2}+\alpha R_{0}e\right],
\end{align}
where
\begin{equation}\label{75}
D_{22}^{(P)}=\frac{2\alpha R_{0}'B_{0}'r^{2}}{B_{0}^{3}}-\frac{2\alpha r}{B_{0}^{2}}\left(R_{0}'+R_{0}''r+\frac{R_{0}'A_{0}'r}{A_{0}}\right).
\end{equation}

\begin{align}\label{76}
Q(r) & =\frac{1}{\left[\frac{2\lambda}{\left(-1+\lambda\right)}\frac{2\left(\bar{c}+\frac{br}{B_{0}}\right)}{B_{0}A_{0}^{2}r^{2}} +\frac{bZ_{R_{0}}R_{0}'}{A_{0}^{2}B_{0}^{2}Z}\right]}\left[\frac{Z_{R_{0}}R_{0}'}{B_{0}Z^{2}}\left(\lambda +\left(1+\lambda\right)\Gamma\frac{p_{r_{0}}}{\rho_{0}+p_{r_{0}}}\right)\frac{\left(-1+\lambda\right)}{2\lambda}\left[\frac{2\alpha e}{Z}\left(\rho_{0}-\frac{R_{0}+\alpha R_{0}^{2}+\lambda T_{0}}{2}\right)\right.\right.\nonumber \\
 & -\frac{2\lambda\left(1+\lambda\right)}{\left(-1+\lambda\right)}\left\{ \frac{b}{B_{0}}\left(p_{r_{0}}+\rho_{0}\right)+\frac{2\bar{c}}{r}\left(\rho_{0}+p_{\perp_{0}}\right)-\frac{e}{4}\right\} \left.+\frac{\lambda e}{2}+A_{0}Z\mathscr{Z}_{1}^{(P)}\right]\nonumber \\
 & -\left[\frac{2\left(1+\lambda\right)}{B_{0}Zr}\left[\Gamma\frac{p_{r_{0}}}{\rho_{0}+p_{r_{0}}} +\frac{\lambda}{1+\lambda}\right]\frac{\left(-1+\lambda\right)}{2\lambda}\left[\frac{2\alpha e}{Z}\left(\rho_{0}-\frac{R_{0}+\alpha R_{0}^{2}+\lambda T_{0}}{2}\right)\right.\right.\nonumber \\
 & -\frac{2\lambda\left(1+\lambda\right)}{\left(-1+\lambda\right)}\left\{ \frac{b}{B_{0}}\left(p_{r_{0}}+\rho_{0}\right)+\frac{2\bar{c}}{r}\left(\rho_{0}+p_{\perp_{0}}\right)-\frac{e}{4}\right\} \nonumber \\
 & \left.+\frac{\lambda e}{2}+A_{0}Z\mathscr{Z}_{1}^{(P)}\right]+\frac{2\left(1+\lambda\right)H_{\perp}}{B_{0}Zr} -\frac{A_{0}'}{A_{0}B_{0}Z}\left(\Gamma\frac{p_{r_{0}}}{\rho_{0}+p_{r_{0}}}+1\right)\frac{\left(-1+\lambda\right)}{2\lambda}\left[\frac{2\alpha e}{Z}\left(\rho_{0}-\frac{R_{0}+\alpha R_{0}^{2}+\lambda T_{0}}{2}\right)\right.\nonumber \\
 & -\frac{2\lambda\left(1+\lambda\right)}{\left(-1+\lambda\right)}\left\{ \frac{b}{B_{0}}\left(p_{r_{0}}+\rho_{0}\right)+\frac{2\bar{c}}{r}\left(\rho_{0}+p_{\perp_{0}}\right)-\frac{e}{4}\right\} \left.+\frac{\lambda e}{2}+A_{0}Z\mathscr{Z}_{1}^{(P)}\right]\nonumber \\
 & -\frac{\left(1+\lambda\right)}{B_{0}Z}\left[\Gamma\frac{p_{r_{0}}}{\rho_{0}+p_{r_{0}}}\frac{\left(-1+\lambda\right)}{2\lambda}\left[\frac{2\alpha e}{Z}\left(\rho_{0}-\frac{R_{0}+\alpha R_{0}^{2}+\lambda T_{0}}{2}\right)-\frac{2\lambda\left(1+\lambda\right)}{\left(-1+\lambda\right)}\left\{ \frac{b}{B_{0}}\left(p_{r_{0}}+\rho_{0}\right)+\frac{2\bar{c}}{r}\left(\rho_{0}+p_{\perp_{0}}\right)-\frac{e}{4}\right\} \right.\right.\nonumber \\
 & \left.\left.+\frac{\lambda e}{2}+A_{0}Z\mathscr{Z}_{1}^{(P)}\right]\right]'-\frac{\lambda}{B_{0}Z}\frac{\left(-1+\lambda\right)}{2\lambda}\left[\frac{2\alpha e}{Z}\left(\rho_{0}-\frac{R_{0}+\alpha R_{0}^{2}+\lambda T_{0}}{2}\right)\right.\nonumber \\
 & \left.\left.-\frac{2\lambda\left(1+\lambda\right)}{\left(-1+\lambda\right)}\left\{ \frac{b}{B_{0}}\left(p_{r_{0}}+\rho_{0}\right)+\frac{2\bar{c}}{r}\left(\rho_{0}+p_{\perp_{0}}\right)-\frac{e}{4}\right\} +\frac{\lambda e}{2}+A_{0}Z\mathscr{Z}_{1}^{(P)}\right]'\right]\frac{2\lambda}{\left(-1+\lambda\right)} -\left[-\frac{Z_{R_{0}}R_{0}'}{B_{0}Z^{2}}\frac{e}{2}\left(Z+\lambda\right)\right.\nonumber \\
 & -\frac{2\alpha e'}{B_{0}Z^{2}}\left(\rho_{0}\lambda+p_{r_{0}}\left(1+\lambda\right)+\frac{R_{0}+\alpha R_{0}^{2}+\lambda T_{0}}{2}\right)+\frac{2\lambda}{\left(-1+\lambda\right)}\left[\frac{2\bar{c}'}{B_{0}Zr}\left(1+\lambda\right)\left(p_{r_{0}} -p_{\perp_{0}}\right)\right.\nonumber \\
 & +\frac{\left(1+\lambda\right)}{A_{0}B_{0}Z}\left[a'\left(p_{r_{0}}+\rho_{0}\right)+ap_{r_{0}}'\right]+\frac{\lambda e'}{2B_{0}Z}-\frac{\alpha eR_{0}'}{B_{0}Z}+\frac{eZ_{R_{0}}R_{0}'}{2B_{0}Z}\nonumber \\
 & \left.\left.\left.+\frac{a}{A_{0}B_{0}Z}\left\{ \rho_{0}'\lambda-\frac{ZR_{0}'}{2}-\frac{\left(R_{0}'+2\alpha R_{0}R_{0}'+\lambda T_{0}'\right)}{2}\right\} \right]+\frac{e'\lambda\left(1+\lambda\right)}{2B_{0}Z\left(-1+\lambda\right)}+\mathscr{Z}_{2}^{(P)}\right]\right].
\end{align}

\begin{align}\label{77}
G(r,t) & =\frac{1}{\left[\frac{2\lambda}{\left(-1+\lambda\right)}\frac{2\left(\bar{c}+\frac{br}{B_{0}}\right)}{B_{0}A_{0}^{2}r^{2}} +\frac{bZ_{R_{0}}R_{0}'}{A_{0}^{2}B_{0}^{2}Z}\right]}\left[-\frac{4\eta\left(1+\lambda\right)}{B_{0}Z} \left\{\frac{2\lambda}{\left(-1+\lambda\right)}\left(\bar{\sigma}'+\frac{\bar{\sigma}A_{0}'}{A_{0}}+\frac{2\bar{\sigma}}{r}\right) -\frac{\bar{\sigma}Z_{R_{0}}R_{0}'}{Z}\right\} +\frac{Z_{R_{0}}R_{0}'}{B_{0}Z^{2}}\left[\bar{\epsilon}\left(1+\lambda\right)\right. \right. \nonumber \\
 & + \left(\lambda+\left(1+\lambda\right)\Gamma\frac{p_{r_{0}}}{\rho_{0} +p_{r_{0}}}\right)\left\{\frac{\left(-1+\lambda\right)}{2\lambda}\frac{A_{0}}{B_{0}}\left[\frac{Z_{R_{0}}R_{0}'}{Z} -\frac{4\lambda}{\left(-1+\lambda\right)}\left(\frac{A_{0}'}{A_{0}}+\frac{1}{r}\right)\right]\left(1+\lambda\right)\int\left(\bar{q} +\bar{\epsilon}\right)dt-\left(1+\lambda\right)\bar{\epsilon}\right.\nonumber \\
 & \left.\left.-\frac{A_{0}}{B_{0}}\left(1+\lambda\right)\int\left(\bar{\epsilon}'+\bar{q}'\right)dt\right\}\right]\nonumber \\
 & -\left[\frac{2\left(1+\lambda\right)}{B_{0}Zr}\left[\Gamma\frac{p_{r_{0}}}{\rho_{0}+p_{r_{0}}} +\frac{\lambda}{1+\lambda}\right]\left[\frac{\left(-1+\lambda\right)}{2\lambda}\frac{A_{0}}{B_{0}}\left[\frac{Z_{R_{0}}R_{0}'}{Z} -\frac{4\lambda}{\left(-1+\lambda\right)}\left(\frac{A_{0}'}{A_{0}}+\frac{1}{r}\right)\right]\left(1+\lambda\right)\int\left(\bar{q} +\bar{\epsilon}\right)dt\right.\right.\nonumber \\
 & \left.-\frac{A_{0}}{B_{0}}\left(1+\lambda\right)\int\left(\bar{\epsilon}'+\bar{q}'\right)dt-\left(1+\lambda\right)\bar{\epsilon}\right] -\frac{\left(1+\lambda\right)\bar{\epsilon}'}{B_{0}Z}-\frac{\left(1+\lambda\right)\left(\dot{\bar{\epsilon}}+\dot{\bar{q}}\right)}{A_{0}Z} -\frac{2\left(1+\lambda\right)\bar{\epsilon}}{B_{0}Zr}-\frac{2\bar{\epsilon}A_{0}'}{A_{0}B_{0}Z}\nonumber \\
 & -\frac{A_{0}'}{A_{0}B_{0}Z}\left(\Gamma\frac{p_{r_{0}}}{\rho_{0}+p_{r_{0}}} +1\right)\left[\frac{\left(-1+\lambda\right)}{2\lambda}\frac{A_{0}}{B_{0}}\left[\frac{Z_{R_{0}}R_{0}'}{Z} -\frac{4\lambda}{\left(-1+\lambda\right)}\left(\frac{A_{0}'}{A_{0}}+\frac{1}{r}\right)\right]\left(1+\lambda\right)\int\left(\bar{q} +\bar{\epsilon}\right)dt\right.\nonumber \\
 & \left.-\frac{A_{0}}{B_{0}}\left(1+\lambda\right)\int\left(\bar{\epsilon}'+\bar{q}'\right)dt-\left(1+\lambda\right)\bar{\epsilon}\right]\nonumber \\
 & -\frac{\left(1+\lambda\right)}{B_{0}Z}\left[\Gamma\frac{p_{r_{0}}}{\rho_{0} +p_{r_{0}}}\left[\frac{\left(-1+\lambda\right)}{2\lambda}\frac{A_{0}}{B_{0}}\left[\frac{Z_{R_{0}}R_{0}'}{Z} -\frac{4\lambda}{\left(-1+\lambda\right)}\left(\frac{A_{0}'}{A_{0}} +\frac{1}{r}\right)\right]\left(1+\lambda\right)\int\left(\bar{q}+\bar{\epsilon}\right)dt\right.\right.\nonumber \\
 & \left.\left.-\left(1+\lambda\right)\bar{\epsilon}-\frac{A_{0}}{B_{0}}\left(1+\lambda\right)\int\left(\bar{\epsilon}'+\bar{q}'\right)dt\right]\right]' \nonumber \\
 & -\frac{\lambda}{B_{0}Z}\left[\frac{\left(-1+\lambda\right)}{2\lambda}\frac{A_{0}}{B_{0}}\left[\frac{Z_{R_{0}}R_{0}'}{Z} -\frac{4\lambda}{\left(-1+\lambda\right)}\left(\frac{A_{0}'}{A_{0}}+\frac{1}{r}\right)\right]\left(1+\lambda\right)\int\left(\bar{q} +\bar{\epsilon}\right)dt\right.\nonumber \\
 & \left.\left.\left.-\left(1+\lambda\right)\bar{\epsilon}-\frac{A_{0}}{B_{0}}\left(1+\lambda\right)\int\left(\bar{\epsilon}' +\bar{q}'\right)dt\right]'\right]\frac{2\lambda}{\left(-1+\lambda\right)}\right].
\end{align}

The explicit form of equation \eqref{63} is obtained in the form
\begin{align}\label{explicit}
\frac{4\eta\left(1+\lambda\right)}{B_{0}Z}\left[\frac{2\lambda}{\left(-1+\lambda\right)}\left(\bar{\sigma}'+\frac{\bar{\sigma}A_{0}'}{A_{0}} +\frac{2\bar{\sigma}}{r}\right)-\frac{\bar{\sigma}Z_{R_{0}}R_{0}'}{Z}\right] -\frac{Z_{R_{0}}R_{0}'}{B_{0}Z^{2}}\left[\bar{\epsilon}\left(1+\lambda\right)+\right.\nonumber \\
\left(\lambda+\left(1+\lambda\right)\Gamma\frac{p_{r_{0}}}{\rho_{0} +p_{r_{0}}}\right)\left[\frac{\left(-1+\lambda\right)}{2\lambda}\frac{A_{0}}{B_{0}}\left[\frac{Z_{R_{0}}R_{0}'}{Z} -\frac{4\lambda}{\left(-1+\lambda\right)}\left(\frac{A_{0}'}{A_{0}} +\frac{1}{r}\right)\right]\left(1+\lambda\right)\int\left(\bar{q}+\bar{\epsilon}\right)dt\right.\nonumber \\
-\left(1+\lambda\right)\bar{\epsilon}-\frac{A_{0}}{B_{0}}\left(1+\lambda\right)\int\left(\bar{\epsilon}'+\bar{q}'\right)dt +D\frac{\left(-1+\lambda\right)}{2\lambda}\left[\frac{2\alpha e}{Z}\left(\rho_{0}-\frac{R_{0}+\alpha R_{0}^{2}+\lambda T_{0}}{2}\right)\right.\nonumber \\
\left.\left.\left.-\frac{2\lambda\left(1+\lambda\right)}{\left(-1+\lambda\right)}\left\{ \frac{b}{B_{0}}\left(p_{r_{0}}+\rho_{0}\right)+\frac{2\bar{c}}{r}\left(\rho_{0}+p_{\perp_{0}}\right)-\frac{e}{4}\right\} +\frac{\lambda e}{2}+A_{0}Z\mathscr{Z}_{1}^{(P)}\right]\right]\right]\nonumber \\
+\frac{2\lambda}{\left(-1+\lambda\right)}\left[\frac{2\left(1+\lambda\right)\bar{\epsilon}}{B_{0}Zr} +\frac{2\left(1+\lambda\right)}{B_{0}Zr}\left[\Gamma\frac{p_{r_{0}}}{\rho_{0}+p_{r_{0}}} +\frac{\lambda}{1+\lambda}\right]\left[\frac{\left(-1+\lambda\right)}{2\lambda}\frac{A_{0}}{B_{0}}\left[\frac{Z_{R_{0}}R_{0}'}{Z}-\right.\right.\right.\nonumber \\
\left.\frac{4\lambda}{\left(-1+\lambda\right)}\left(\frac{A_{0}'}{A_{0}}+\frac{1}{r}\right)\right]\left(1+\lambda\right)\int\left(\bar{q} +\bar{\epsilon}\right)dt-\frac{A_{0}}{B_{0}}\left(1+\lambda\right)\int\left(\bar{\epsilon}'+\bar{q}'\right)dt-\left(1+\lambda\right)\bar{\epsilon}\nonumber \\
+D\frac{\left(-1+\lambda\right)}{2\lambda}\left[\frac{2\alpha e}{Z}\left(\rho_{0}-\frac{R_{0}+\alpha R_{0}^{2}+\lambda T_{0}}{2}\right)-\frac{2\lambda\left(1+\lambda\right)}{\left(-1+\lambda\right)}\left\{ \frac{b}{B_{0}}\left(p_{r_{0}}+\rho_{0}\right)+\frac{2\bar{c}}{r}\left(\rho_{0}+p_{\perp_{0}}\right)-\frac{e}{4}\right\} \right.\nonumber \\
\left.\left.+\frac{\lambda e}{2}+A_{0}Z\mathscr{Z}_{1}^{(P)}\right]\right]-\frac{2\left(1+\lambda\right)DH_{\perp}}{B_{0}Zr}+\frac{2\ddot{D}\left(\bar{c} +\frac{br}{B_{0}}\right)}{B_{0}A_{0}^{2}r^{2}}+\frac{2\bar{\epsilon}A_{0}'}{A_{0}B_{0}Z}\nonumber \\
+\frac{A_{0}'}{A_{0}B_{0}Z}\left(\Gamma\frac{p_{r_{0}}}{\rho_{0}+p_{r_{0}}} +1\right)\left[\frac{\left(-1+\lambda\right)}{2\lambda}\frac{A_{0}}{B_{0}}\left[\frac{Z_{R_{0}}R_{0}'}{Z} -\frac{4\lambda}{\left(-1+\lambda\right)}\left(\frac{A_{0}'}{A_{0}}+\frac{1}{r}\right)\right]\left(1+\lambda\right)\int\left(\bar{q} +\bar{\epsilon}\right)dt\right.\nonumber \\
-\frac{A_{0}}{B_{0}}\left(1+\lambda\right)\int\left(\bar{\epsilon}'+\bar{q}'\right)dt-\left(1+\lambda\right)\bar{\epsilon}\nonumber \\
+D\frac{\left(-1+\lambda\right)}{2\lambda}\left[\frac{2\alpha e}{Z}\left(\rho_{0}-\frac{R_{0}+\alpha R_{0}^{2}+\lambda T_{0}}{2}\right)-\frac{2\lambda\left(1+\lambda\right)}{\left(-1+\lambda\right)}\left\{ \frac{b}{B_{0}}\left(p_{r_{0}}+\rho_{0}\right)+\frac{2\bar{c}}{r}\left(\rho_{0}+p_{\perp_{0}}\right)-\frac{e}{4}\right\} \right.\nonumber \\
\left.\left.+\frac{\lambda e}{2}+A_{0}Z\mathscr{Z}_{1}^{(P)}\right]\right]+\frac{\left(1+\lambda\right)\bar{\epsilon}'}{B_{0}Z}+\frac{\left(1+\lambda\right)\left(\dot{\bar{\epsilon}} +\dot{\bar{q}}\right)}{A_{0}Z}\nonumber \\
+\frac{\left(1+\lambda\right)}{B_{0}Z}\left[\Gamma\frac{p_{r_{0}}}{\rho_{0} +p_{r_{0}}}\left[\frac{\left(-1+\lambda\right)}{2\lambda}\frac{A_{0}}{B_{0}}\left[\frac{Z_{R_{0}}R_{0}'}{Z} -\frac{4\lambda}{\left(-1+\lambda\right)}\left(\frac{A_{0}'}{A_{0}}+\frac{1}{r}\right)\right]\left(1+\lambda\right)\int\left(\bar{q} +\bar{\epsilon}\right)dt\right.\right.\nonumber \\
-\left(1+\lambda\right)\bar{\epsilon}-\frac{A_{0}}{B_{0}}\left(1+\lambda\right)\int\left(\bar{\epsilon}'+\bar{q}'\right)dt\nonumber \\
+D\frac{\left(-1+\lambda\right)}{2\lambda}\left[\frac{2\alpha e}{Z}\left(\rho_{0}-\frac{R_{0}+\alpha R_{0}^{2}+\lambda T_{0}}{2}\right)-\frac{2\lambda\left(1+\lambda\right)}{\left(-1+\lambda\right)}\left\{ \frac{b}{B_{0}}\left(p_{r_{0}}+\rho_{0}\right)+\frac{2\bar{c}}{r}\left(\rho_{0}+p_{\perp_{0}}\right)-\frac{e}{4}\right\} \right.\nonumber \\
\left.\left.\left.+\frac{\lambda e}{2}+A_{0}Z\mathscr{Z}_{1}^{(P)}\right]\right]\right]' +\frac{\lambda}{B_{0}Z}\left[\frac{\left(-1+\lambda\right)}{2\lambda}\frac{A_{0}}{B_{0}}\left[\frac{Z_{R_{0}}R_{0}'}{Z} -\frac{4\lambda}{\left(-1+\lambda\right)}\left(\frac{A_{0}'}{A_{0}} +\frac{1}{r}\right)\right]\left(1+\lambda\right)\int\left(\bar{q}+\bar{\epsilon}\right)dt\right.\nonumber \\
-\left(1+\lambda\right)\bar{\epsilon}-\frac{A_{0}}{B_{0}}\left(1+\lambda\right)\int\left(\bar{\epsilon}'+\bar{q}'\right)dt\nonumber \\
+D\frac{\left(-1+\lambda\right)}{2\lambda}\left[\frac{2\alpha e}{Z}\left(\rho_{0}-\frac{R_{0}+\alpha R_{0}^{2}+\lambda T_{0}}{2}\right)-\frac{2\lambda\left(1+\lambda\right)}{\left(-1+\lambda\right)}\left\{ \frac{b}{B_{0}}\left(p_{r_{0}}+\rho_{0}\right)+\frac{2\bar{c}}{r}\left(\rho_{0}+p_{\perp_{0}}\right)-\frac{e}{4}\right\} \right.\nonumber \\
\left.\left.\left.+\frac{\lambda e}{2}+A_{0}Z\mathscr{Z}_{1}^{(P)}\right]\right]'\right]+\frac{\ddot{D}bZ_{R_{0}}R_{0}'}{A_{0}^{2}B_{0}^{2}Z} +D\left[-\frac{Z_{R_{0}}R_{0}'}{B_{0}Z^{2}}\frac{e}{2}\left(Z+\lambda\right)-\frac{2\alpha e'}{B_{0}Z^{2}}\left(\rho_{0}\lambda+p_{r_{0}}\left(1+\lambda\right)+\frac{R_{0}+\alpha R_{0}^{2}+\lambda T_{0}}{2}\right)\right.\nonumber \\
+\frac{2\lambda}{\left(-1+\lambda\right)}\left[\frac{2\bar{c}'}{B_{0}Zr}\left(1+\lambda\right)\left(p_{r_{0}}-p_{\perp_{0}}\right)\right. +\frac{\left(1+\lambda\right)}{A_{0}B_{0}Z}\left[a'\left(p_{r_{0}}+\rho_{0}\right)+ap_{r_{0}}'\right]+\frac{\lambda e'}{2B_{0}Z}-\frac{\alpha eR_{0}'}{B_{0}Z}+\frac{eZ_{R_{0}}R_{0}'}{2B_{0}Z}\nonumber \\
\left.\left.+\frac{a}{A_{0}B_{0}Z}\left\{ \rho_{0}'\lambda-\frac{ZR_{0}'}{2}-\frac{\left(R_{0}'+2\alpha R_{0}R_{0}'+\lambda T_{0}'\right)}{2}\right\} \right]+\frac{e'\lambda\left(1+\lambda\right)}{2B_{0}Z\left(-1+\lambda\right)}+\mathscr{Z}_{2}^{(P)}\right]=0.
\end{align}

\begin{align}\label{78}
H_{1} & =-\frac{4\eta\left(1+\lambda\right)}{Z}\left[\frac{2\lambda}{\left(-1+\lambda\right)}\left(\bar{\sigma}'+\frac{2\bar{\sigma}}{r}\right) -\frac{\bar{\sigma}Z_{R_{0}}R_{0}'}{Z}\right]+\frac{Z_{R_{0}}R_{0}'}{Z^{2}}\left[\bar{\epsilon}\left(1+\lambda\right)+\lambda\bar{\rho}_{N}\right]\nonumber \\
 & +\frac{2\lambda}{\left(-1+\lambda\right)}\left[-\frac{2\bar{\epsilon}\left(1+\lambda\right)}{Zr}\right.-\frac{2\lambda\bar{\rho}_{N}}{Zr} +\frac{2\left(1+\lambda\right)H_{\perp_{N}}}{Zr}\frac{1}{2\sqrt{Q}}\left[e^{\sqrt{Q}t}\int Ge^{-\sqrt{Q}t}dt-e^{-\sqrt{Q}t}\int Ge^{\sqrt{Q}t}dt\right]\nonumber \\
 & -\left[G+\frac{\sqrt{Q}}{2}\left[e^{\sqrt{Q}t}\int Ge^{-\sqrt{Q}t}dt-e^{-\sqrt{Q}t}\int Ge^{\sqrt{Q}t}dt\right]\right]\left[\frac{2\left(\bar{c}+br\right)}{r^{2}}+\frac{\left(-1+\lambda\right)}{2\lambda}\frac{bZ_{R_{0}}R_{0}'}{Z}\right]\nonumber \\
 & \left.-\frac{\left(1+\lambda\right)}{Z}\left[\bar{\epsilon}'+\dot{\bar{\epsilon}}+\dot{\bar{q}}\right]-\frac{\lambda\bar{\rho}_{N}'}{Z}\right]\nonumber \\
 & -\frac{1}{2\sqrt{Q}}\left[e^{\sqrt{Q}t}\int Ge^{-\sqrt{Q}t}dt-e^{-\sqrt{Q}t}\int Ge^{\sqrt{Q}t}dt\right]\left[-\frac{Z_{R_{0}}R_{0}'e\left(Z+\lambda\right)}{2Z^{2}}\right.\nonumber \\
 & -\frac{2\alpha e'}{Z^{2}}\left(\lambda\rho_{0}+\frac{R_{0}+\alpha R_{0}^{2}+\lambda T_{0}}{2}\right)\nonumber \\
 & +\frac{2\lambda}{\left(-1+\lambda\right)}\left[\frac{2\bar{c}'}{Zr}\left(1+\lambda\right)\left(p_{r_{0}}-p_{\perp_{0}}\right) +\frac{a'\left(1+\lambda\right)\rho_{0}}{Z}+\frac{\lambda e'}{2Z}-\frac{\alpha eR_{0}'}{Z}\right.\nonumber \\
 & \left.\left.+\frac{eZ_{R_{0}}R_{0}'}{2Z}+\frac{a}{Z}\left\{ \rho_{0}'\lambda-\frac{ZR_{0}'}{2}-\frac{\left(R_{0}'+2\alpha R_{0}R_{0}'+\lambda T_{0}'\right)}{2}\right\} +\mathscr{Z}_{2_{N}}^{(P)}\right]\right].
\end{align}

\begin{align}\label{79}
H_{2} & =-\frac{4\eta r\left(1+\lambda\right)}{\left(r+m_{0}\right)Z}\left[\frac{2\lambda}{\left(-1+\lambda\right)}\left(\bar{\sigma}' +\frac{m_{0}\bar{\sigma}}{r\left(r-m_{0}\right)}+\frac{2\bar{\sigma}}{r}\right)-\frac{\bar{\sigma}Z_{R_{0}}R_{0}'}{Z}\right]\nonumber \\
 & +\frac{Z_{R_{0}}R_{0}'r}{\left(r+m_{0}\right)Z^{2}}\left[\bar{\epsilon}\left(1+\lambda\right)+\lambda\bar{\rho}_{pN}\right] +\frac{2\lambda}{\left(-1+\lambda\right)}\left[-\frac{2\bar{\epsilon}\left(1+\lambda\right)}{\left(r+m_{0}\right)Z}\right.\nonumber \\
 & -\frac{2\lambda\bar{\rho}_{pN}}{\left(r+m_{0}\right)Z} +\frac{2\left(1+\lambda\right)H_{\perp_{pN}}}{\left(r+m_{0}\right)Z}\frac{1}{2\sqrt{Q}}\left[e^{\sqrt{Q}t}\int Ge^{-\sqrt{Q}t}dt-e^{-\sqrt{Q}t}\int Ge^{\sqrt{Q}t}dt\right]\nonumber \\
 & -\left[G+\frac{\sqrt{Q}}{2}\left[e^{\sqrt{Q}t}\int Ge^{-\sqrt{Q}t}dt-e^{-\sqrt{Q}t}\int Ge^{\sqrt{Q}t}dt\right]\right]\left(\frac{2r\left(\bar{c}+\frac{br^{2}}{r+m_{0}}\right)}{\left(r+m_{0}\right)\left(r-m_{0}\right)^{2}} +\frac{\left(-1+\lambda\right)}{2\lambda}\frac{bZ_{R_{0}}R_{0}'r^{4}}{\left(r^{2}-m_{0}^{2}\right)^{2}Z}\right)\nonumber \\
 & -\frac{\left(1+\lambda\right)r}{Z}\left[\frac{\bar{\epsilon}'}{\left(r+m_{0}\right)}+\frac{\dot{\bar{\epsilon}}+\dot{\bar{q}}}{\left(r-m_{0}\right)}\right] -\frac{m_{0}}{\left(r^{2}-m_{0}^{2}\right)Z}\left(2\bar{\epsilon}+\bar{\rho}_{pN}\right)-\frac{\lambda r\bar{\rho}_{pN}'}{\left(r+m_{0}\right)Z}\nonumber \\
 & -\frac{1}{2\sqrt{Q}}\left[e^{\sqrt{Q}t}\int Ge^{-\sqrt{Q}t}dt-e^{-\sqrt{Q}t}\int Ge^{\sqrt{Q}t}dt\right]\left[-\frac{Z_{R_{0}}R_{0}'re\left(Z+\lambda\right)}{2Z^{2}\left(r+m_{0}\right)}\right.\nonumber \\
 & -\frac{2\alpha e'r}{\left(r+m_{0}\right)Z^{2}}\left(\lambda\rho_{0}+p_{r_{0}}\left(1+\lambda\right)+\frac{R_{0}+\alpha R_{0}^{2}+\lambda T_{0}}{2}\right)+\frac{2\lambda}{\left(-1+\lambda\right)}\left[\frac{2\bar{c}'}{Z\left(r+m_{0}\right)}\left(1+\lambda\right)\left(p_{r_{0}} -p_{\perp_{0}}\right)\right.\nonumber \\
 & +\frac{a'r^{2}\left(1+\lambda\right)\left(p_{r_{0}}+\rho_{0}\right)}{\left(r^{2}-m_{0}^{2}\right)Z}+\frac{\lambda e'r}{2Z\left(r+m_{0}\right)}-\frac{\alpha erR_{0}'}{Z\left(r+m_{0}\right)}+\frac{erZ_{R_{0}}R_{0}'}{2\left(r+m_{0}\right)Z}\nonumber \\
 & \left.\left.+\frac{ar^{2}}{Z\left(r^{2}-m_{0}^{2}\right)}\left\{ \rho_{0}'\lambda-\frac{ZR_{0}'}{2}-\frac{\left(R_{0}'+2\alpha R_{0}R_{0}'+\lambda T_{0}'\right)}{2}\right\} \right]+\frac{e'\lambda\left(1+\lambda\right)r}{2\left(-1+\lambda\right)\left(r+m_{0}\right)Z}+\mathscr{Z}_{2_{pN}}^{(P)}\right].
\end{align}


\begin{thebibliography}{}

\bibitem{Chandra} S. Chandrasekhar, Mon. Not. R. Astr. Soc. \textbf{91}, 456 (1931); ibid. \textbf{95}, 207 (1935); Astrophys. J. \textbf{74}, 81 (1931); Observatory \textbf{57}, 373 (1934).

\bibitem{8} C. Hansen and S. Kawaler, \textit{Stellar Interiors: Physical Principles, Structure and Evolution} (Springer Verlag, 1994).

\bibitem{9} R. Kippenhahn and A. Weigert, \textit{Stellar Structure and Evolution} (Springer Verlag, 1990).

\bibitem{5} E.N. Glass, Phys. Lett. A \textbf{86}, 351 (1981).

\bibitem{Santos} N. O. Santos, Mon. Not. R. Astron. Soc. \textbf{216}, 403 (1985).

\bibitem{OSK} A. K. G. De Oliveira, N. O. Santtos and C. A. Kolassis, Mon. Not. R. Astron. Soc. \textbf{216}, 1001 (1985).

\bibitem{OPS} A. K. G. de Oliveira, J. A. de F. Pacheco, and N. O. Santos, Mon. Not. R. Astron. Soc. \textbf{220}, 405 (1986).

\bibitem{10} L. Herrera, G. Le Denmat and N.O. Santos, Mon. Not. R. Astron. Soc. \textbf{237}, 257 (1989).

\bibitem{14} R. Chan, S. Kichenassamy, G. Le Denmat and N.O. Santos, Mon. Not. R. Astron. Soc. \textbf{239}, 91 (1989).

\bibitem{15} R. Chan, L. Herrera and N.O. Santos, Mon. Not. R. Astron. Soc. \textbf{265}, 533 (1993).

\bibitem{16} R. Chan, L. Herrera and N.O. Santos, Mon. Not. R. Astron. Soc. \textbf{267}, 637 (1994).

\bibitem{11} R. Chan, Mon. Not. R. Astron. Soc. \textbf{316}, 588 (2000).

\bibitem{12} L. Herrera and N.O. Santos, Phys. Rev. D \textbf{70}, 084004 (2004).

\bibitem{13} L. Herrera, N.O. Santos and G. Le Denmat, Gen. Relativ. Gravit. \textbf{44}, 1143 (2012).

\bibitem{GGM} M. Govender, K. S. Govinder and S. D. Maharaj, Int. Jour. Mod. Phys. D \textbf{12}, 667 (2003).

\bibitem{MGG1} S. D. Maharaj, G. Govender and M. Govender, Gen. Relativ. Gravit. \textbf{44}, 1089 (2012).

\bibitem{GMM} M. Govender, R. Maartens and S. D. Maharaj, Mon. Not. R. Astron. Soc. \textbf{310}, 557 (1999).

\bibitem{MGG2} S. D. Maharaj, G. Govender and M. Govender, PRAMANA \textbf{77}, 469 (2011).

\bibitem{GB} S. Guha and R. Banerji, Int. J. Theor. Phys. \textbf{53}, 2332 (2014).

\bibitem{PHDMS} A. Di Prisco, L. Herrera, G. Le Denmat, M. A. H. MacCallum and N. O. Santos, Phys. Rev. D \textbf{76}, 064017 (2007).

\bibitem{17a} S. Nojiri and S. D. Odintsov, Int. J. Geom. Meth. Mod. Phys. \textbf{4}, 115 (2007); Phys. Rept. \textbf{505}, 59 (2011);
    S. Capozziello and M. De Laurentis, Phys. Rept. \textbf{509}, 167 (2011); T. Clifton, P. G. Ferreira, A. Padilla and
    C. Skordis, Phys. Rept. \textbf{513}, 1 (2012);
    O. Bertolami and M. C. Sequeira, Phys. Rev. D \textbf{79}, 104010 (2009);
        T. Harko, F. S. N. Lobo, G. Otalora and E. N. Saridakis, Phys. Rev. D \textbf{89}, 124036 (2014);
        T. Harko, F. S. N. Lobo, G. Otalorac and E.N. Saridakis, JCAP \textbf{12}, 021 (2014);
    M. Sharif and M. Zubair, J. High. Energy Phys. \textbf{12}, 079 (2013).
\bibitem{DEinfl} S. Nojiri and S. Odintsov, Phys. Rev. D \textbf{74}, 086005 (2006); Gen. Rel. Grav. \textbf{36}, 1765 (2004); S. Capozziello, Int. J. Mod. Phys. D \textbf{11}, 483 (2002);
    S. M. Carroll, et al., Phys. Rev. D \textbf{71}, 063513 (2005);
    S. Carloni, P. K. S. Dunsby, S. Capozziello and A. Troisi, Class. Quant. Grav. \textbf{22}, 4839 (2005);
    J. A. R. Cembranos, Phys. Rev. D \textbf{73}, 064029 (2006);
    T. Clifton and J. D. Barrow, Phys. Rev. D \textbf{72}, 103005 (2005);
    K. Bamba, C. Q. Geng and C. C. Lee, JCAP \textbf{08}, 021 (2010).
\bibitem{evolution} A. de la Cruz-Dombriz and A. Dobado, Phys. Rev. D \textbf{74}, 087501 (2006);
    P. K. S. Dunsby, E. Elizalde, R. Goswami, S. Odintsov and D. S\'{a}ez-G\'{o}mez, Phys. Rev. D \textbf{82}, 023519 (2010);
    S. Nojiri and S. D. Odintsov, Phys. Rev. D \textbf{68}, 123512 (2003).

\bibitem{18} S. G. Ghosh and S.D. Maharaj, Phys. Rev. D \textbf{85}, 124064 (2012).

\bibitem{19} J. A. R. Cembranos, A. de la Cruz-Dombriz and B. M. Nunez, JCAP \textbf{04}, 021 (2012).

\bibitem{Goswami} R. Goswami, A. M. Nzioki, S. D. Maharaj and S. G. Ghosh, Phys. Rev. D \textbf{90}, 084011 (2014).

\bibitem{CB} S. Chakrabarti and N. Banerjee, Gen. Relativ. Gravit., \textbf{48}, 57 (2016).

\bibitem{SY2} M. Sharif and Z. Yousaf, Phys. Rev. D \textbf{88}, 024020 (2013).

\bibitem{SY3}M. Sharif and Z. Yousaf, Mon. Not. R. Astron. Soc. \textbf{434}, 2529-2538 (2013).

\bibitem{22} T. Harko, F. S. N. Lobo, S. Nojiri and S.D. Odinstov, Phys. Rev. D \textbf{84}, 024020 (2011).

\bibitem{22b} T. Harko and F. S. N. Lobo, Galaxies \textbf{2}, 410 (2014).

\bibitem{22a} T. Harko and F. S. N. Lobo, Eur. Phys. J. C \textbf{70}, 373 (2010).

\bibitem{22c} T. Harko, Phys. Rev. D \textbf{81}, 044021 (2010).

\bibitem{Starobinsky} A. A. Starobinsky, Phys. Lett. B \textbf{91}, 99 (1980).

\bibitem{SZ2} M. Sharif and M. Zubair, J. Phys. Soc. Jpn. \textbf{81}, 114005 (2012).

\bibitem{alvarenga2013} F. G. Alvarenga, A. de la Cruz-Dombriz, M. J. S. Houndjo, M. E. Rodrigues and D. S\'{a}ez-G\'{o}mez, Phys. Rev. D \textbf{87}, 103526 (2013).

\bibitem{barrientos2014} J. O. Barrientos and G.F. Rubilar, Phys. Rev. D \textbf{90}, 028501 (2014).

\bibitem{Faraoni2009} V. Faraoni, Phys. Rev. D \textbf{80}, 124040 (2009).

\bibitem{23} H. Shabani and M. Farhoudi, Phys. Rev. D. \textbf{90}, 044031 (2014).

\bibitem{24} M. Sharif and M. Zubair, JCAP \textbf{03}, 028 (2012);
    M. Jamil, D. Momeni and R. Myrzakulov, Chin. Phys. Lett. \textbf{29}, 109801 (2012).

\bibitem{SZ1} M. Sharif and M. Zubair, Gen. Relativ. Gravit., \textbf{46}, 1723 (2014).
\bibitem{Houndjo} M. J. S. Houndjo, Int. J. Mod. Phys. D \textbf{21}, 1250003 (2012).
\bibitem{OSG} S. D. Odintsov and D. S\'{a}ez-G\'{o}mez, Phys. Lett. B \textbf{725}, 437 (2013).

\bibitem{SY} M. Sharif and Z. Yousaf, Astrophys. Space Sci. \textbf{354}, 471 (2014).

\bibitem{NZ1} Ifra Noureen and M. Zubair, Astrophys. Space Sci. \textbf{356}, 103 (2015).

\bibitem{NZ2} Ifra Noureen and M. Zubair, Eur. Phys. J. C \textbf{75}, 62 (2015).

\bibitem{NZ3} Ifra Noureen, M. Zubair, A. A. Bhatti and G. Abbas, Eur. Phys. J. C \textbf{75}, 323 (2015).

\bibitem{YBB} Z. Yousaf, K. Bamba and M. Z. Bhatti, Phys. Rev. D \textbf{93}, 124048 (2016).
\bibitem{InflModGrav} Z. Yousaf, K. Bamba and M. Z. Bhatti, Phys. Rev. D \textbf{93}, 064059 (2016).
\bibitem{YBBG} Z. Yousaf, K. Bamba, M. Z. Bhatti and U. Ghafoor, Phys. Rev. D \textbf{100}, 024062 (2019).
\bibitem{Y2020} Z. Yousaf, Phys. Dark Univ. \textbf{28}, 100509 (2020).
\bibitem{BYR} M. Z. Bhatti, Z. Yousaf and A. Rehman, Phys. Dark Univ. \textbf{29}, 100561 (2020).

\bibitem{SS} M. Sharif and A. Siddiqa, Eur. Phys. J. Plus \textbf{132}, 529 (2017).

\bibitem{SS1} M. Sharif and A. Siddiqa, AHEP \textbf{(2019)}, 8702795 (2019).

\bibitem{hdimPfcoll} S. Khan, M. S. Khan and A. Ali, Modern Physics Letters A \textbf{33}, 1850065 (2018); M. Sharif and A. Anwar, Astrophys. Space Sci. \textbf{363}, 123 (2018).

\bibitem{SW} M. Sharif and A. Waseem, Gen Relativ Gravit \textbf{50}, 78 (2018).

\bibitem{AA1} G. Abbas and R. Ahmed, Eur. Phys. J. C \textbf{77}, 441 (2017).

\bibitem{AA} G. Abbas and R. Ahmed, Mod. Phys. Lett. A \textbf{34}, 1950153 (2019).

\bibitem{AA2} R. Ahmed and G. Abbas, Can. J. Phys. \textbf{97}, 994 (2019).

\bibitem{ZA} M. Zubair and H. Azmat, Phys. Dark Univ. \textbf{28}, 100531 (2020).
\bibitem{YBA} Z. Yousaf, M. Z. Bhatti and H. Asad, Phys. Dark Univ. \textbf{28}, 100527 (2020).
\bibitem{YBF} Z. Yousaf, M. Z. Bhatti and U. Farwa, Class. Quant. Grav. \textbf{34}, 145002 (2017).
\bibitem{YBF2} Z. Yousaf, M. Z. Bhatti and U. Farwa, Mon. Not. R. Astron. Soc. \textbf{464}, 45094519 (2017).
\bibitem{YBBF} Z. Yousaf, K. Bamba, M. Z. Bhatti and U. Farwa, Eur. Phys. J. A \textbf{54}, 122 (2018).
\bibitem{Yousaf2018} Z. Yousaf, Astrophys. Space Sci. \textbf{363}, 226 (2018).
\bibitem{Yousaf2019a} Z. Yousaf, Eur. Phys. J. Plus \textbf{134}, 245 (2019).
\bibitem{Yousaf2019b} Z. Yousaf, Mod. Phys. Lett. A \textbf{34}, 1950333 (2019).
\bibitem{Yousaf2021} Z. Yousaf, Eur. Phys. J. Plus \textbf{136}, 281 (2021).

\bibitem{MS} C. W. Misner and D. Sharp, Phys. Rev. \textbf{136} B571 (1964).

\bibitem{Misner} C.W. Misner, Phys. Rev. \textbf{137}, B1360, (1965).

\bibitem{Tolman} R. Tolman, Phys. Rev. \textbf{35}, 904 (1939).

\bibitem{Eckart} C. Eckart, Phys. Rev. \textbf{58}, 919 (1940).

\bibitem{Landau} L. Landau and E. Lifshitz, Fluid Mechanics (Pergamon
Press, London, 1959).

\bibitem{Muller} I. Muller, Z. Physik \textbf{198}, 329 (1967).

\bibitem{Israel1} W. Israel, Ann. Phys. (N.Y.) \textbf{100}, 310 (1976).

\bibitem{Israel2} W. Israel and J. Stewart, Phys. Lett. A \textbf{58}, 213
(1976).

\bibitem{Israel3} W. Israel and J. Stewart, Ann. Phys. (N.Y.) \textbf{118},
341 (1979).

\bibitem{HerreraInertia} L. Herrera, Int. J. Mod. Phys. D \textbf{15}, 2197 (2006).

\bibitem{HM} L. Herrera and J. Martinez, Astrophys. Space Sci. \textbf{259}, 235 (1998).

\bibitem{HDSW} L. Herrera, G. Le. Denmat, N. O. Santos and A. Wang, Int. J. Mod. Phys. D \textbf{13}, 583 (2004).

\bibitem{HPFT2009} L. Herrera, A. D. Prisco, E. Fuenmayor and O. Troconis, Int. Jour. of Mod. Phys. D \textbf{18}, 129, (2009).

\bibitem{Maartens} R. Maartens, Causal Thermodynamics in Relativity. Lectures given at the Hanno Rund Workshop on
Relativity and Thermodynamics, University of Natal, June 1996. astro-ph/9609119.

\bibitem{srGRG10} M. Sharif and Z. Rehmat, Gen Relativ Gravit \textbf{42}, 1795 (2010).

\bibitem{CC} S. Chakraborty and S. Chakraborty, AHEP \textbf{(2017)}, 8786791 (2017).

\bibitem{SA} S. M. Shah and G. Abbas, Eur. Phys. J. C \textbf{77}, 251 (2017).
\bibitem{PS} J. M. Z. Pretel and M. F. A. da Silva, Gen. Relativ. Gravit. \textbf{51}, 3 (2019).

\bibitem{SK} M. Sharif and H. R. Kausar, Journal of Physics: Conference Series \textbf{354}, 012020 (2012).

\bibitem{AbbasNazar2018} G. Abbas and H. Nazar, Adv. High Energy Phys. \textbf{2018}, 9250786 (2018).
\bibitem{PJRA} J. M. Z. Pretel, S. E. Jor\'{a}s, R. R. R. Reis and J. D. V. Arba\~{n}il, arXiv:2012.03342v1 [gr-qc].

\bibitem{AA3} R. Ahmed and G. Abbas, Chinese Journal of Physics \textbf{65}, 177 (2020).


\bibitem{TRM} S. Thirukkanesh, S. S. Rajah and S. D. Maharaj, J. Math. Phys. \textbf{53}, 032506 (2012).

\bibitem{Lake} These calculations were done with the help of Maple program and GRTensor package: K. Lake and P. J. Musgrave,
    GRTensor (Queen's University, Kingston, 2003).

\bibitem{C2} S. Chandrasekhar, Astrophys. J. \textbf{140}, 417 (1964).

\bibitem{HS} L. Herrera and N.O. Santos, Phys. Rept. \textbf{286}, 53 (1997).

\bibitem{HW} B.K. Harrison, K.S. Thorne, J.A. Wheeler, \emph{Gravitation Theory and Gravitational Collapse }(University of Chicago Press, Chicago,
1965).

\end{thebibliography}
\end{document}